%% file: main.tex
\documentclass[lettersize,journal]{IEEEtran}
\usepackage{amsmath,amsfonts}
\usepackage{algorithmic}
\usepackage{algorithm}
\usepackage{array}
\usepackage[caption=false,font=normalsize,labelfont=sf,textfont=sf]{subfig}
\usepackage{textcomp}
\usepackage{stfloats}
\usepackage{url}
\usepackage{verbatim}
\usepackage{graphicx}
\usepackage{cite}
\usepackage{xspace,xcolor,colortbl}
\usepackage{hyperref}
\usepackage{multirow}
\usepackage[final]{microtype}
\usepackage{enumitem}

\usepackage{booktabs}
\usepackage{siunitx}    
\usepackage{wrapfig}

\usepackage{booktabs,tabularx,makecell,xcolor,caption,subcaption,adjustbox}
\newcolumntype{Y}{>{\centering\arraybackslash}X}
\definecolor{resultpurple}{HTML}{7B1FA2} 
\setlist[itemize]{noitemsep, topsep=0pt, leftmargin=*}
\begin{document}

\title{Cross-Domain Acceleration of Open Modification Search: From Commodity Platforms to Emerging Memory and Storage Devices}


\author{
Sumukh Pinge$^*$,~\IEEEmembership{Graduate Student Member,~IEEE}, 
Chang Eun Song$^*$,~\IEEEmembership{Graduate Student Member,~IEEE}, \\
Po-Kai Hsu,~\IEEEmembership{Graduate Student Member,~IEEE},
Zheyu Li,
Ashkan Moradifirouzabadi,~\IEEEmembership{Graduate Student Member,~IEEE},
Yanru Chen,
Xiangjin Wu,~\IEEEmembership{Graduate Student Member,~IEEE},
Wei-Chen Chen,\\
Eric Pop,~\IEEEmembership{Fellow,~IEEE},
Shimeng Yu,~\IEEEmembership{Fellow,~IEEE},
H.-S. Philip Wong,~\IEEEmembership{Life Fellow,~IEEE},\\
Tajana Rosing,~\IEEEmembership{Fellow,~IEEE}, and 
Mingu Kang,~\IEEEmembership{Member,~IEEE}
\thanks{$^*$The authors contribute equally.}
\thanks{Sumukh Pinge, Chang Eun Song, Zheyu Li, and Tajana Rosing are with the Department of Computer Science and Engineering, University of California San Diego, La Jolla, CA 92093 USA.}
\thanks{Yanru Chen, Ashkan Moradifirouzabadi, and Mingu Kang are with the Department of Electrical and Computer Engineering, University of California San Diego, La Jolla, CA 92093 USA.}
\thanks{Po-Kai Hsu and Shimeng Yu are with the School of Electrical and Computer Engineering, Georgia Institute of Technology, Atlanta, GA 30332-0250 USA.}
\thanks{Xiangjin Wu, Wei-Chen Chen, Eric Pop, and H.-S. Philip Wong are with the Department of Electrical Engineering, Stanford University, Stanford, CA 94305-9505 USA.}
\thanks{Manuscript received XXX; revised XXX.}
}


\markboth{IEEE Journal on Emerging and Selected Topics in Circuits and Systems}%
{Pinge \MakeLowercase{\textit{et al.}}: Cross-Domain Acceleration of Open Modification Search}

\maketitle

\begin{abstract}
Open modification search (OMS) in mass spectrometry (MS) is a data-intensive workload whose performance is dominantly limited by reference data movement rather than computation. Prior OMS accelerators have largely been evaluated in isolation, making it difficult to understand system-level trade-offs across platforms.
This paper presents the first workload-driven, cross-platform survey of accelerators for MS search by studying not only commodity platforms, but also emerging memory- and storage-centric architectures, including GPUs, near-storage FPGAs, DRAM near-memory processing, ReRAM/PCM in-memory processing, and 3D NAND/FeNAND in-storage processing, under consistent algorithmic and accuracy assumptions. Leveraging a binary hyperdimensional computing (HDC)–based OMS formulation that reduces similarity evaluation to lightweight bitwise primitives and tolerates device-level non-idealities, we enable a robust execution on memory-centric architectures despite device-level non-idealities and limited computing capability. Overall, this study identifies memory- and storage-centric architectures as a key architectural breakthrough for large-scale, high-speed search acceleration, delivering up to $>$$100\times$ speedup and $>$$40,000\times$ improvement in energy efficiency. 

\end{abstract}


\input{TEX/1.Introduction}

\input{TEX/2.Background}

\input{TEX/3_4_5.Architectures_Accelerators_Mapping.tex}

\input{TEX/5.OMS_Model}

\input{TEX/6.MergedEval.tex}
\input{TEX/7.Discussions}

\input{TEX/8.Conclusions}

\raggedbottom
\bibliographystyle{IEEEtran}
\bibliography{JETCAS}

\newlength{\biophotowidth}
\setlength{\biophotowidth}{0.78in}

\newcommand{\compactbio}[3]{%
  \par\addvspace{0.45\baselineskip}
  \begingroup
  \footnotesize
  \setlength{\intextsep}{0pt}%
  \setlength{\columnsep}{0.08in}
  \begin{wrapfigure}[7]{l}{\biophotowidth}
    \vspace{-1\baselineskip}
    \includegraphics[width=\biophotowidth,height=0.98in,clip,keepaspectratio]{#1}
    \vspace{0.1\baselineskip}
  \end{wrapfigure}%
  \noindent\textbf{#2} #3\par
  \endgroup
}

\vspace*{4mm}
\compactbio{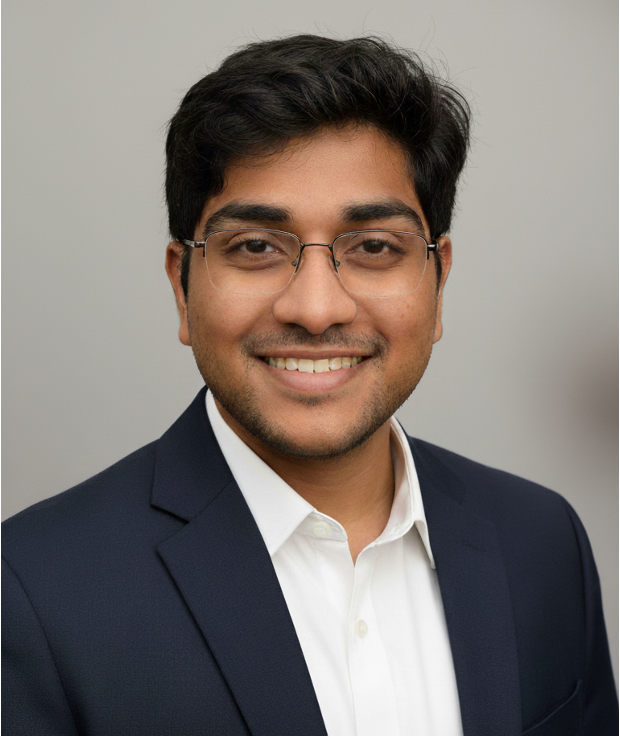}{Sumukh Pinge}{%
(Graduate Student Member, IEEE) received the B.E. degree in Electronics and Instrumentation from Birla Institute of Technology and Science, Pilani, India, in 2022, and the Ph.D. degree in Computer Science from the University of California San Diego, La Jolla, CA, USA, in 2026. His research interests include Hardware--Software Co-Design with a focus on Memory- and Storage-Centric Similarity Search Accelerators, FPGA Systems, Hyperdimensional Computing, Proteomics Acceleration, and Retrieval-Augmented Generation (RAG).
}

\vspace*{4mm}

\compactbio{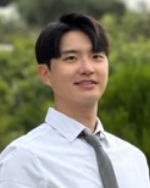}{Chang Eun (Paul) Song}{(Graduate Student Member, IEEE) received the B.E. degree in Electrical Engineering from Korea University, Seoul, South Korea, in 2022. He is currently pursuing the Ph.D. degree in Computer Science and Engineering at the University of California, San Diego, La Jolla, CA, USA. His research interests include HW/SW co-design for AI accelerators, digital and mixed-signal chip design with a processing-in-memory (PIM), and optimization for large language models (LLMs), retrieval-augmented generation (RAG), and neuromorphic computing. He was a recipient of the UCSD Best Doctoral Research Award in 2025 and the Samsung Electronics Global Talent Ph.D. Fellowship in 2023.
}
\vspace*{3mm}

\compactbio{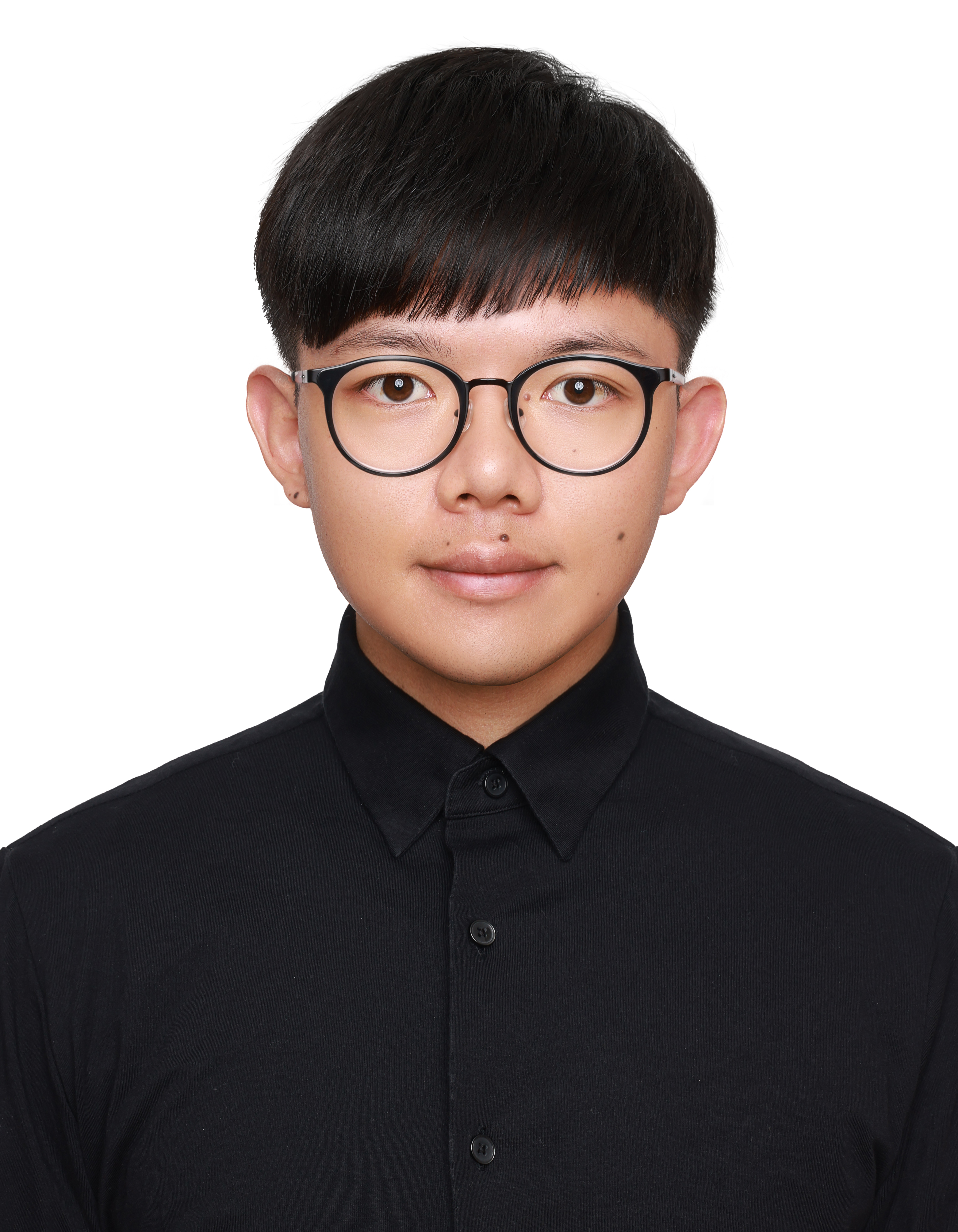}{Po-Kai Hsu}{%
(Graduate Student Member, IEEE) received the B.S. degree in electrical engineering and the M.S. degree in electro-optical engineering from National Tsing Hua University, Hsinchu, Taiwan, in 2014 and 2018, respectively, and the Ph.D. degree in electrical and computer engineering from the Georgia Institute of Technology, Atlanta, GA, USA, in 2025. From 2018 to 2021, he was a Device Engineer with the Emerging Central Laboratory, Macronix International Company Ltd., Hsinchu. His research interests include algorithm-hardware co-design and system-technology co-optimization for hardware accelerators, with a focus on genome sequencing applications and large language models.
}

\vspace*{3mm}
\compactbio{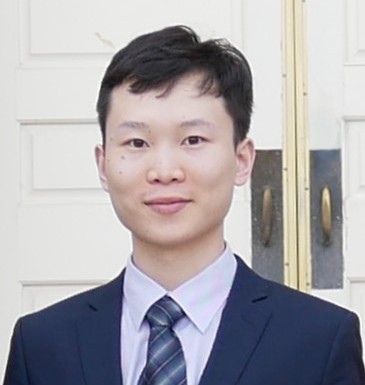}{Zheyu Li}{%
received the B.Sc. degree in Electrical Engineering and the Ph.D. degree in Computer Science and Engineering from Pennsylvania State University, University Park, PA, USA, in 2017 and 2023, respectively. He currently is a postdoc in the Energy Efficiency Lab at the University of California, San Diego (UCSD). His research interests lie in hardware and software acceleration of data-intensive emerging workloads, including bioinformatics applications, brain-inspired computing, near-memory computing, and FPGA-based architectures.
}

\vspace*{2mm}
\compactbio{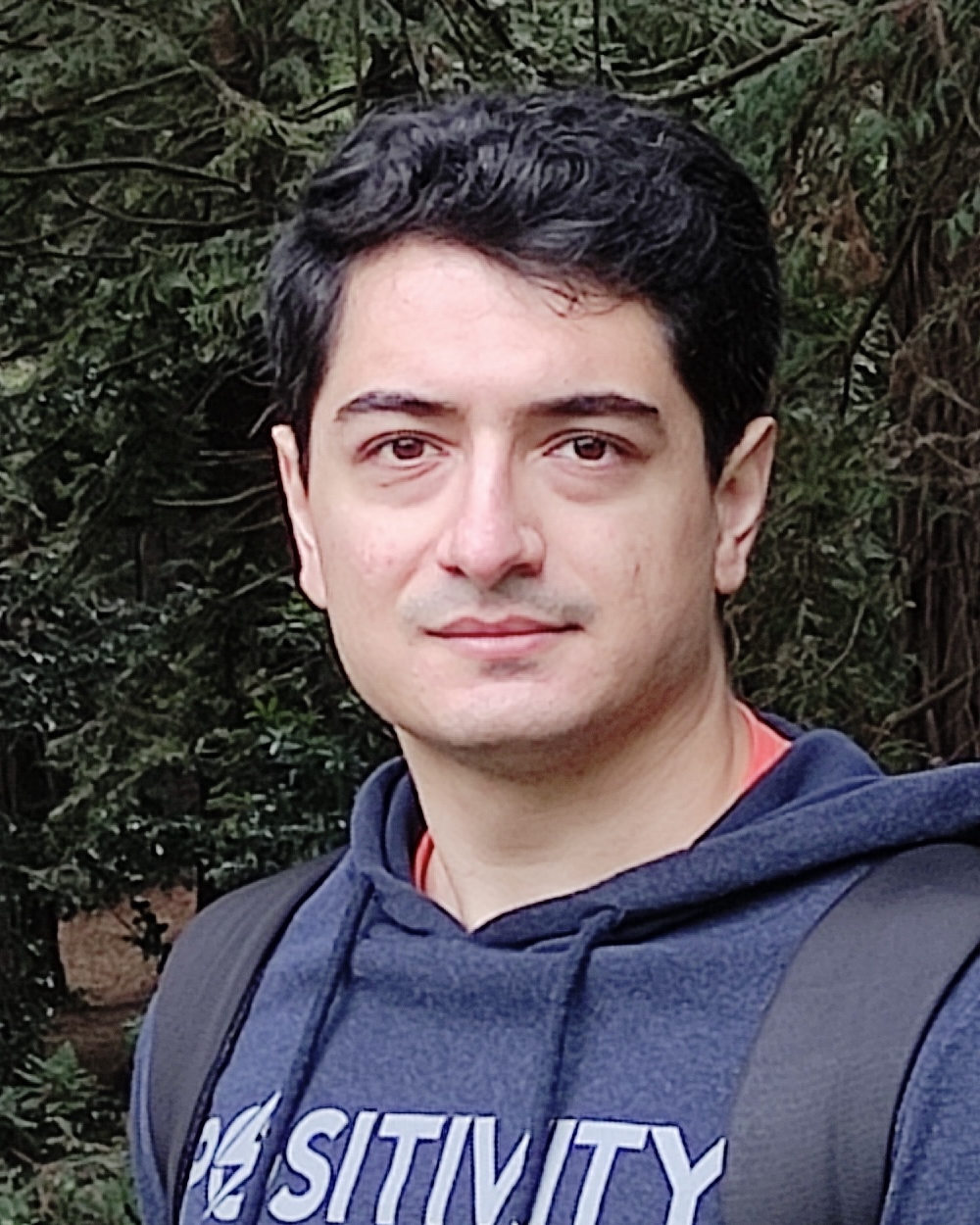}{Ashkan Moradifirouzabadi}{%
(Graduate Student Member, IEEE) received the B.S. degree in electrical engineering from the University of Tehran, Tehran, Iran, in 2021 and the M.S. degree in computer engineering from the University of California, San Diego, CA, USA, in 2024. He is currently pursuing a Ph.D. degree in computer engineering at the University of California, San Diego, CA, USA. His research interests include hardware-algorithm co-design for machine learning, machine learning systems, and in-memory computing architectures.
}

\vspace*{4mm}
\compactbio{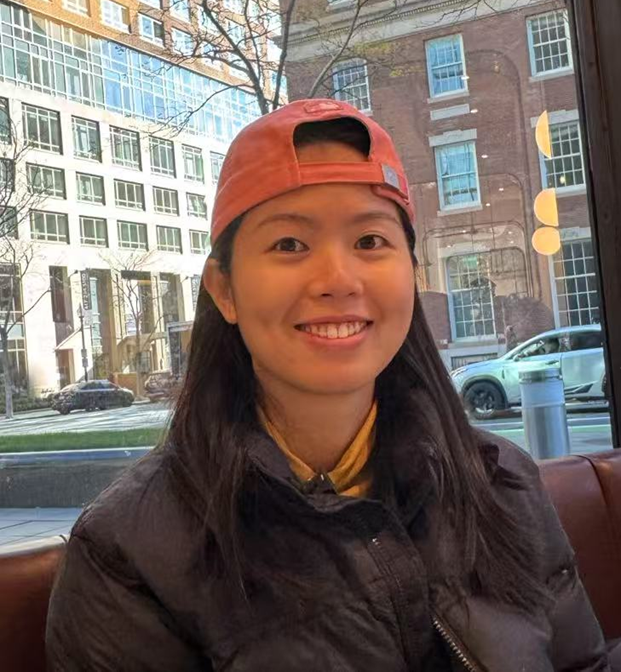}{Yanru Chen}{is currently a Ph.D. candidate in Electrical and Computer Engineering at the University of California San Diego, La Jolla, CA, USA. She received the M.S. degree in Electronic Information (Intelligent Manufacturing) from Tsinghua University, Beijing, China, in 2024, and the B.E. degree in Electronic Science and Technology from Jilin University, Changchun, China, in 2021. Her research interests include SW/HW co-design, processing-in-memory, in-storage processing, graph analytics, hyperdimensional computing, and database search.
}

\vspace*{4mm}
\compactbio{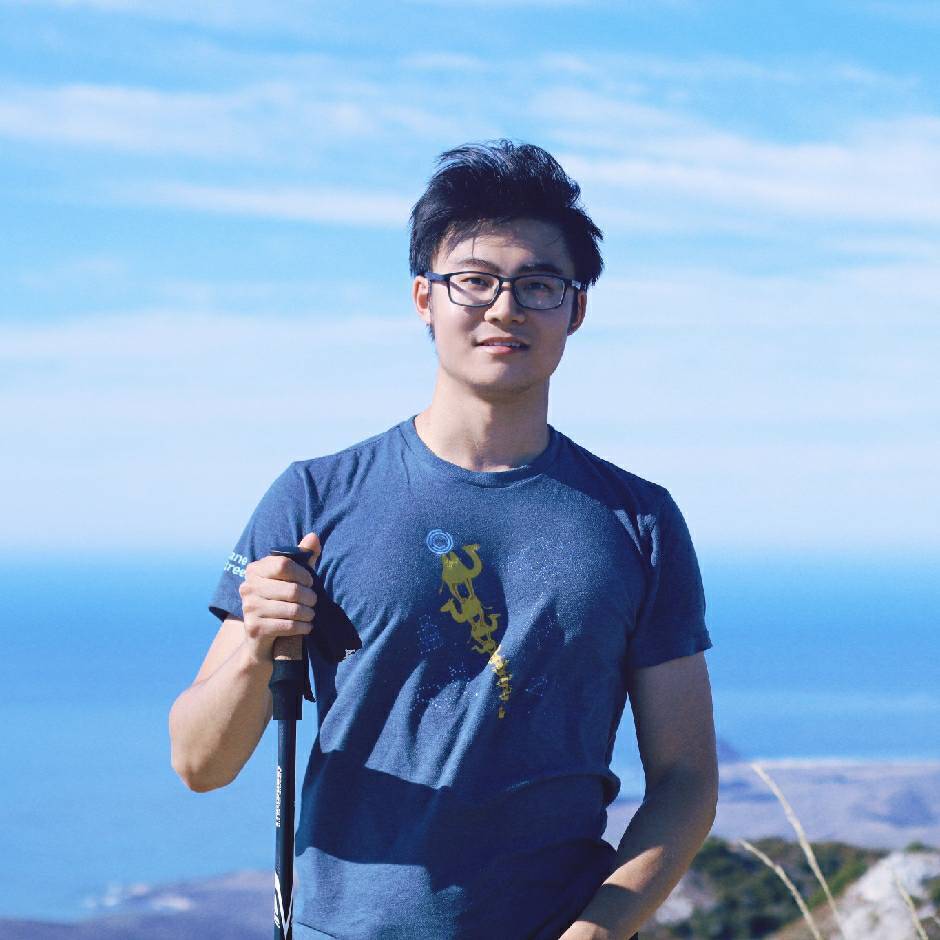}{Xiangjin Wu}{received his B.S. in Physics from Nanjing University (2020) and his M.S. and Ph.D. in EE at Stanford (2025). His research focuses on novel materials and heterostructures for memory applications, including phase change memory (PCM), dynamic random-access memory (DRAM), and their interconnects.
}

\vspace*{4mm}
\compactbio{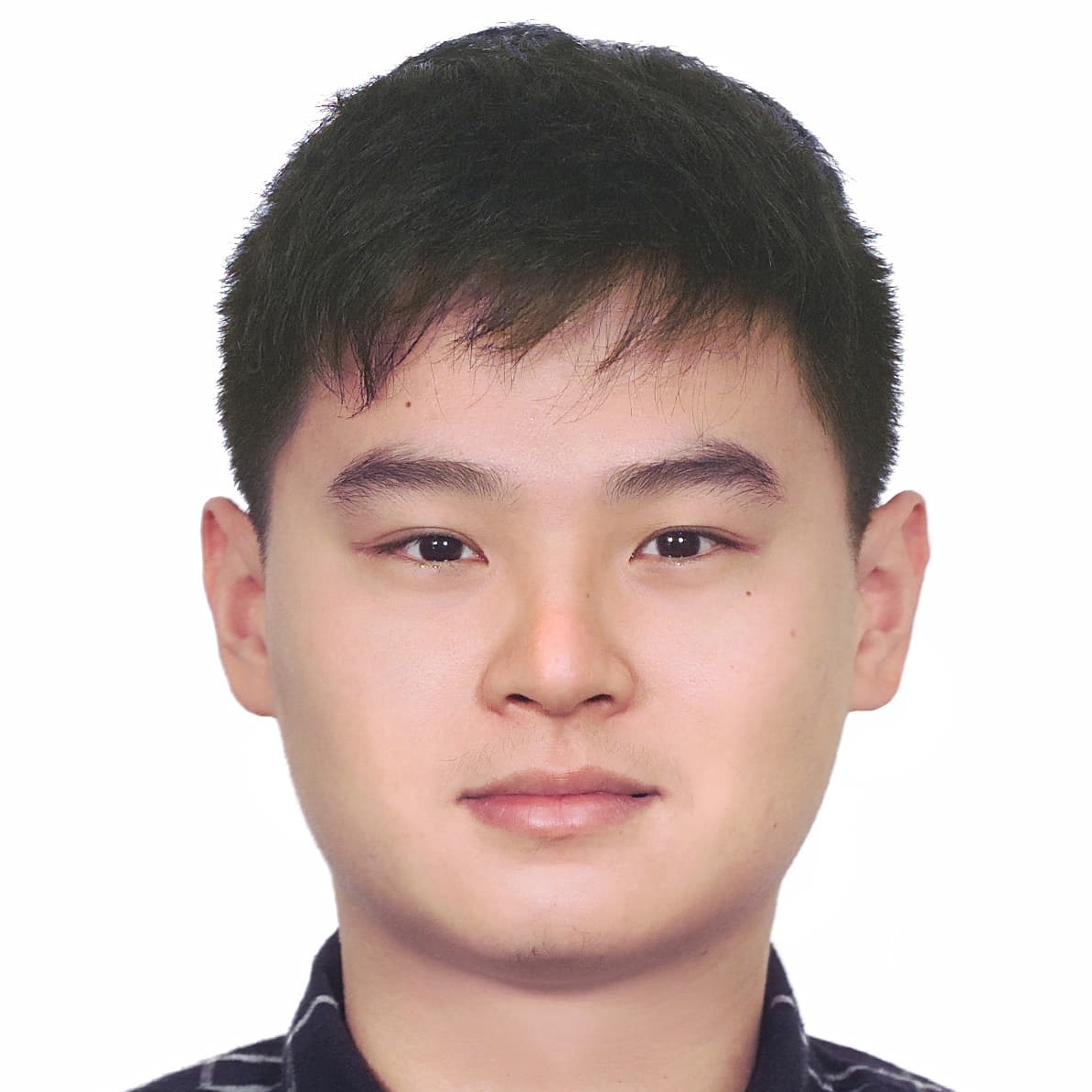}{Wei-Chen Chen}{received the Ph.D. degree in electrical engineering from Stanford University, Stanford, CA, USA. His research interests include in-memory computing hardware and its associated algorithm design.
}
\vspace*{9mm}
\compactbio{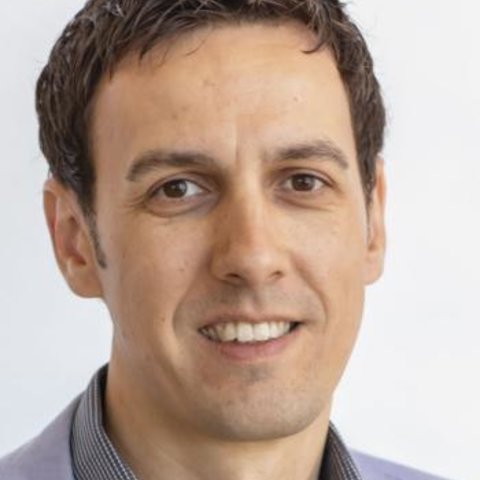}{Eric Pop}{%
(Fellow, IEEE) is the Pease-Ye Professor of Electrical Engineering and, by courtesy, of Materials Science and Engineering and Applied Physics at Stanford University and SLAC. He is also the faculty co-lead of the SystemX Alliance and a Senior Fellow at the Precourt Institute for Energy. Prior to Stanford, he was on the faculty of the University of Illinois at Urbana-Champaign and worked in industry at Intel and IBM. He received the Ph.D. degree in Electrical Engineering from Stanford University, Stanford, CA, USA, in 2005, and the M.Eng. and B.S. degrees in Electrical Engineering and Computer Science and the B.S. degree in Physics from the Massachusetts Institute of Technology, Cambridge, MA, USA, in 1999. His research interests include semiconductors, nanoelectronics, data storage, and energy. His honors include the Intel Outstanding Researcher Award, the Presidential Early Career Award for Scientists and Engineers (PECASE), Young Investigator Awards from the Navy, Air Force, NSF, and DARPA, and several best-paper awards with his students. He is an APS Fellow and a Clarivate Highly Cited Researcher.
}
\vspace*{4mm}

\compactbio{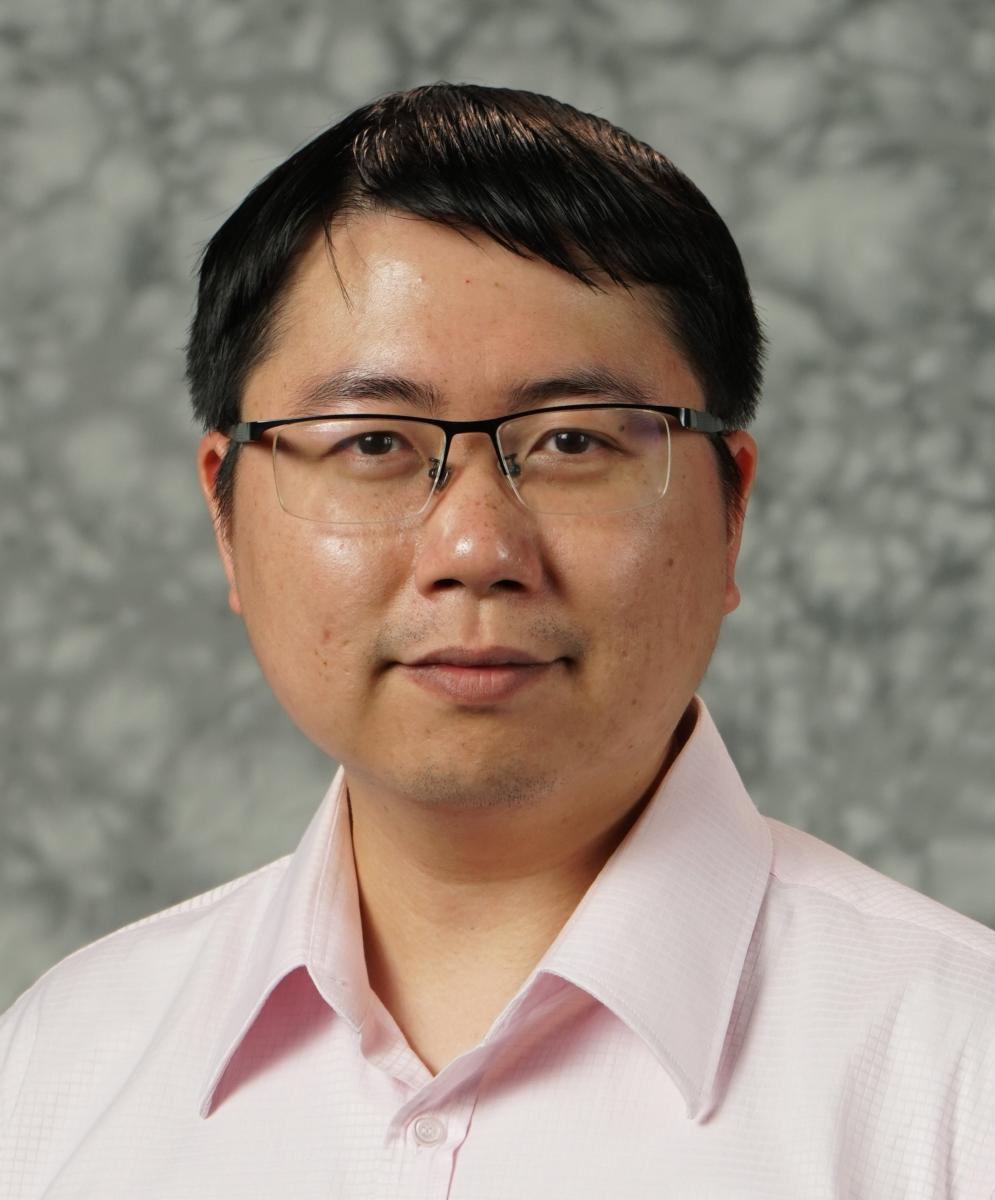}{Shimeng Yu}{%
(Fellow, IEEE) is a Full Professor of Electrical and Computer Engineering with the Georgia Institute of Technology, Atlanta, GA, USA, where he holds the Dean's Professorship. He received the B.S. degree in Microelectronics from Peking University, Beijing, China, in 2009, and the M.S. and Ph.D. degrees in Electrical Engineering from Stanford University, Stanford, CA, USA, in 2011 and 2013, respectively. From 2013 to 2018, he was an Assistant Professor with Arizona State University. His research interests include semiconductor devices and integrated circuits for energy-efficient computing systems, with expertise in emerging non-volatile memories for AI hardware and 3-D integration. He was elevated to IEEE Fellow for contributions to non-volatile memories and in-memory computing. He was a recipient of the NSF Faculty Early Career Award in 2016, the IEEE Electron Devices Society Early Career Award in 2017, the ACM SIGDA Outstanding New Faculty Award in 2018, the Semiconductor Research Corporation Young Faculty Award in 2019, the ACM/IEEE Design Automation Conference Under-40 Innovators Award in 2020, and the Intel Outstanding Researcher Award in 2023.}

\compactbio{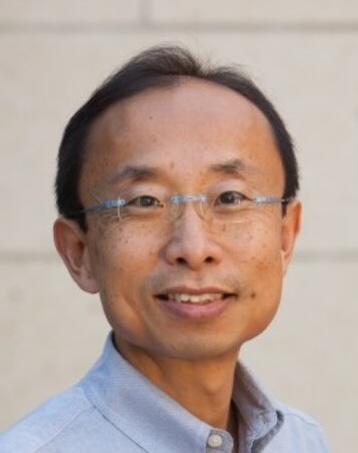}{H.-S. Philip Wong}{%
(Fellow, IEEE) is the Willard R. and Inez Kerr Bell Professor in the School of Engineering and a Professor of Electrical Engineering at Stanford University, Stanford, CA, USA. He joined Stanford University in 2004 after working at the IBM T. J. Watson Research Center from 1988 to 2004. From 2018 to 2020, he was on leave from Stanford University and served as Vice President of Corporate Research at Taiwan Semiconductor Manufacturing Company (TSMC), where he has served as Chief Scientist in a consulting and advisory role since 2020. He is the founding Faculty Co-Director of the Stanford SystemX Alliance and the Faculty Director of the Stanford Nanofabrication Facility. His research interests include semiconductor devices, nanotechnology, memory technologies, and microelectronics. He received the IEEE Electron Devices Society J. J. Ebers Award, the IEEE Andrew S. Grove Award, and the IEEE Technical Field Award for outstanding contributions to solid-state devices and technology.
}

\vspace*{4mm}

\compactbio{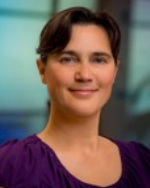}{Tajana Rosing}{%
(Fellow, IEEE) is a Professor, Holder of the Fratamico Endowed Chair, and Director of the System Energy Efficiency Laboratory, University of California at San Diego, La Jolla, CA, USA. She received the Ph.D. degree from Stanford University, Stanford, CA, USA, in 2001. From 1998 to 2005, she was a full-time Research Scientist with HP Labs, Palo Alto, CA, USA, while also leading research efforts with Stanford University, Stanford, CA, USA. She was a Senior Design Engineer with Altera Corporation, San Jose, CA, USA. She is leading a number of projects, including efforts funded by the DARPA/SRC JUMP 2.0 PRISM program with a focus on the design of accelerators for analysis of big data, DARPA- and NSF-funded projects on hyperdimensional computing, and an SRC-funded project on IoT system reliability and maintainability. Her current research interests include energy-efficient computing, cyber--physical, and distributed systems.
}

\vspace*{4mm}

\compactbio{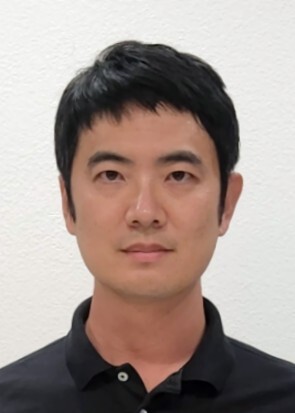}{Mingu Kang}{%
(Member, IEEE) is an Assistant Professor of Electrical and Computer Engineering (ECE) with the Jacobs School of Engineering, University of California at San Diego, La Jolla, CA, USA. He received the B.S. and M.S. degrees in Electrical and Electronic Engineering from Yonsei University, Seoul, South Korea, in 2007 and 2009, respectively, and the Ph.D. degree in Electrical and Computer Engineering from the University of Illinois at Urbana-Champaign, Champaign, IL, USA, in 2017. From 2009 to 2012, he was with the Memory Division, Samsung Electronics, Hwaseong, South Korea, where he was involved in the circuit and architecture design of phase change memory (PRAM). From 2017 to 2020, he was a Research Staff Member of the IBM Thomas J. Watson Research Center, Yorktown Heights, NY, USA, where he designed machine-learning accelerator architecture, which was successfully commercialized by being embedded on IBM Z-mainframe server. Dr. Kang was a recipient of Intel 2022 Rising Star Award, 2024 Hellman Fellow Award, the UIUC CSL Best Thesis Award in 2018, the MICRO TOP Pick Honorable Mention 2019, and the IEEE International Symposium on Circuits and Systems (ISCAS) ``Neural System and Application'' Best Paper Awards in 2016 and 2018.

}

\end{document}

%% file: TEX/1.Introduction.tex
\section{Introduction}
\IEEEPARstart{M}{ass} spectrometry (MS) has become a cornerstone in proteomics and metabolomics, enabling high-throughput measurement of peptides,  proteins, and their post-translational modifications (PTMs). Each tandem MS (MS/MS) experiment can generate millions of fragmentation spectra that must be matched against a reference database or spectral library. As repositories grow into the terabyte (TB) scale \cite{massive2024, martens2005pride}, the database search step has become the dominant bottleneck. In particular, open modification search (OMS) expands the search space by allowing wide precursor mass tolerances to capture unexpected PTMs, thereby enabling novel biological insights\cite{Duarte2016}. While OMS substantially improves identification coverage, it also entails billions of comparisons per dataset, creating severe demands on both computation and data movement~\cite{annsolo}.

To address this challenge, a wide range of accelerators have been proposed, with commodity platforms such as GPUs and FPGAs exploiting massive parallelism. In particular, GPUs (e.g., NVIDIA RTX~4090 and H100) provide thousands of CUDA cores and device-memory bandwidth in the TB-per-second range ($\sim$1TB/s GDDR6X on RTX~4090; $\sim$3.35TB/s HBM on H100)~\cite{nvidiaH100whitepaper}. FPGAs, by contrast, enable customized data paths and can be integrated near memory or storage to partially reduce data movement~\cite{lee2020smartssd}. Nonetheless, both architectures maintain a clear separation between memory and compute, which fundamentally limits their efficiency on highly data-intensive workloads where data transfer dominates execution time~\cite{wulf1995memorywall,gebregiorgis2022memorycentric}.

New memory-centric architectural paradigms have emerged as promising solutions to this limitation. These include DRAM, Resistive RAM (ReRAM), and phase-change memory (PCM) based In-Memory Processing (IMP) and Near-Memory Processing (NMP) accelerators, as well as 3D NAND and ferroelectric NAND (FeNAND) based In-Storage Processing (ISP) devices, which colocate compute with data to minimize data movement \cite{ambit17,upmem23,neurram22,pcmreview24,smartssd21,ngd2021,lee2021_3dnand}. In parallel, algorithmic reformulations based on \emph{hyperdimensional computing} (HDC) encode spectra into binary (BIN) hypervectors (HVs) and reduce similarity evaluation to simple XOR and population count (popcount) operations \cite{kanerva2009,imani2017voicehd}. Such lightweight primitives are well-suited for IMP and NMP platforms, which often support limited computational complexity \cite{ambit17,simdram21,upmem23}. Collectively, these approaches demonstrate substantial speedups and energy savings for OMS workloads \cite{kang2023homstc,kang2022massively,hsu2023memsys,fan2024mlcrram,fan2024specpcm,hsu2025fenand,pinge2024rapidoms,kang2024dram}.

Despite a wide spectrum of studies shown above, the field still lacks systematic cross-device comparisons that clarify relative strengths and weaknesses of each approach. Individual works often evaluate accelerators in isolation, using different datasets, accuracy targets, and hardware assumptions, which makes it difficult to assess true trade-offs across platforms. For instance, GPUs deliver strong performance on moderate datasets but fail to scale to multi-TB libraries; DRAM provides fast access but remains capacity-limited; ReRAM and PCM provide massive throughput gains but face endurance and drift challenges; and 3D NAND/FeNAND offer unmatched density yet suffer from sequential access and limited computing efficiency, unlike crossbar-type devices.

This paper narrows this gap by presenting a workload-driven comparative survey of DRAM\cite{kang2024dram}, ReRAM\cite{fan2024mlcrram}, PCM\cite{fan2024specpcm}, 3D NAND\cite{hsu2023memsys}, and FeNAND\cite{hsu2025fenand}-based accelerators for OMS, along with GPU\cite{kang2023homstc,kang2022massively} and FPGA\cite{pinge2024rapidoms} based references. This contrasts with prior studies that examined devices in isolation or focused only on device-level properties, without connecting them to end-to-end workload behavior. While perfect fairness for the comparison is challenging due to inherent differences in device characteristics, capacity, and operating mechanisms, we aim to provide as consistent basis as possible. 
To ensure fairness, we unify assumptions across platforms. All accelerators are evaluated using the same reference database, under practical device memory capacity constraints, and with equivalent accuracy requirements. We explore platform-specific architectural choices together with data mapping strategies, highlighting trade-offs in speed, energy efficiency, scalability, and workload suitability.
Specifically, we evaluate all devices using a mid-scale dataset, and additionally study large-scale datasets for storage-level 3D NAND and FeNAND devices. By providing the first cross-platform perspective on OMS acceleration, this work aims to guide device, architecture, and algorithm co-optimization for the next generation of mass spectrometry accelerators. While this paper focuses on OMS, the insights extend to a broader class of search-centric and data-intensive applications beyond proteomics.

The remainder of this paper is organized as follows: Section \ref{sec:background} provides background on OMS dataflow and HDC algorithm. Section \ref{sec:arch-mapping} presents various accelerators based on commodity processors and emerging memory technologies. Section \ref{sec:eval} and \ref{Results} evaluate these approaches under a unified methodology. 



%% file: TEX/2.Background.tex
\section{Background}\label{sec:background}
\begin{figure}[t]
    \centering
        \centering
        \includegraphics[width=1\linewidth]{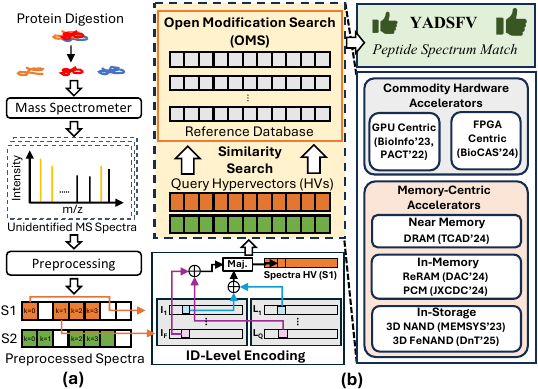}
        \caption{End-to-end OMS pipeline: (a) preprocessing and (b) HDC implementation and their accelerations.}
    \label{fig:oms}
\end{figure}
\subsection{OMS search in tandem mass spectrometry}
Mass spectrometry identifies molecules based on their mass-to-charge ratio ($m/z$)~\cite{zhang1998universal}. In proteomics, proteins are enzymatically digested into peptides; in tandem mass spectrometry (MS/MS), a precursor peptide is isolated, fragmented, and the resulting fragment ions are measured across $m/z$. An MS spectrum is an x-y graph: intensity on the y-axis versus  $m/z$ on the x-axis as shown in Fig.~\ref{fig:oms}(a). Spectral library search identifies an unknown spectrum by comparing it against a library of reference spectra obtained from known peptides.

After precursor isolation, MS/MS fragments the precursor and measures the resulting fragment-ion masses, and the resulting spectrum is preprocessed to select informative peaks via peak filtering, noise reduction, and normalization. 
To enable library search, we then construct a vector representation of the spectrum by discretizing the $m/z$ axis via binning, which improves search quality but increases the memory footprint for sparse representations. We consider two common choices for this representation: (1) binned sparse vectors, whose similarity is evaluated using shifted cosine similarity to tolerate small $m/z$ misalignments~\cite{annsolo}; 
and (2) HD-encoded vectors whose similarity is computed either with efficient XOR and popcount (detailed in  Section~\ref{sec:hdc}) or with cosine similarity. 


A standard pass uses a narrow precursor window and metadata filters based on the precursor  pair $(m/z,\ \text{charge})$, which is typically sufficient for accurate identification of unmodified peptides. However, post-translational modifications shift precursor mass and alter fragment patterns\cite{creasy2004ptm}, so a modified query may not closely match any single unmodified reference. To address this, open modification search (OMS) is introduced \cite{annsolo}, which widens the precursor mass window so that modified queries can be matched against largely unmodified libraries. This improves identification coverage in complex or previously unexplored samples, but also enlarges the candidate set per query, increasing compute and data-movement cost. At the repository scale, libraries contain millions of references and modern experiments generate massive volumes of spectra. 
As a result, the search stage becomes the dominant contributor to both runtime and energy, since each query must be compared against a substantial portion of the library.
The implications of such full scan type of operations on data movement, data preparation cost, and device capacity are discussed in Section~\ref{Results}.
Finally, the scorer ranks candidates by similarity and passes the top-$k$ candidates to a peptide--spectrum match (PSM) validation stage that controls false discoveries via false discovery rate (FDR) filtering (e.g., target--decoy FDR)~\cite{elias2007target}.

\subsection{Hyperdimensional computing for OMS} \label{sec:hdc}

As shown in Fig.~\ref{fig:oms}(b), HDC encodes each spectrum into a $D$-bit binary HV, allowing similarity computation to be carried out using simple bitwise operations. Given a preprocessed peak list with binned $m/z$ indices $b_k$ and quantized intensity levels $\ell_k$, HDC maps each peak to a high-dimensional representation through two codebooks: ID hypervectors ${\mathbf{I}_i \in \{0,1\}^D}$ for $F$ discretized $m/z$ bins and level hypervectors ${\mathbf{L}_j\in{\{0,1\}}^D}$ for $Q$ intensity levels~\cite{morris2022HyDREA}, where the codebooks $\{\mathbf{I}_i\}$ and $\{\mathbf{L}_j\}$ are generated offline once and shared across spectra. The methods used to generate the codebooks are described in the following paragraph. Here, $b_k$ and $\ell_k$ are discrete symbols used as indices into these codebooks, i.e.,  the $k$-th peak selects its position HV as $\mathbf{I}_{b_k}$ and its intensity HV as $\mathbf{L}_{\ell_k}$. Each peak is then encoded by binding its ID and level HVs using XOR,
\begin{equation}
\mathbf{u}_k = \mathbf{I}_{b_k} \oplus \mathbf{L}_{\ell_k},
\label{eq:peak-bind}
\end{equation}
The final spectrum HV is obtained via component-wise majority voting over the $P$ peaks in the spectrum, i.e., $\{\mathbf{u}_k\}_{k=1}^{P}$:
\begin{equation}
\mathbf{h} = \operatorname{Majority}(\mathbf{u}_1,\ldots,\mathbf{u}_P),
\label{eq:majority}
\end{equation}
where for each bit position $d \in \{1,\ldots,D\}$, $\mathbf{h}[d]=1$ if $\sum_{k=1}^{P}\mathbf{u}_k[d] > P/2$ and $\mathbf{h}[d]=0$ otherwise. 

To enhance robustness against peak shifts and intensity variations, we employ locality-preserving codebooks. The first ID hypervector $\mathbf{I}_1$ is initialized randomly, and each subsequent $\mathbf{I}_i$ is generated by flipping a small, fixed number of bits from $\mathbf{I}_{i-1}$ so that adjacent $m/z$ bins exhibit controlled Hamming distance. Level hypervectors ${\mathbf{L}_j}$ are constructed similarly by flipping $D/Q$ bits between adjacent intensity levels. With this method, small perturbations in peak location or intensity modify only a limited fraction of bits in each bound vector of \eqref{eq:peak-bind}, and the majority operation in \eqref{eq:majority} naturally suppresses such localized noise, producing stable HVs suitable for OMS~\cite{kang2022massively,imani2017voicehd}.
Finally, the similarity between query and reference spectra, represented by $\mathbf{h}_q$ and $\mathbf{h}_r$, is computed using normalized Hamming distance:
\begin{equation}
s_H(\mathbf{h}_q,\mathbf{h}_r)
= 1 - \frac{\mathrm{popcount}(\mathbf{h}_q \oplus \mathbf{h}_r)}{D},
\label{eq:hamming}
\end{equation}
where $\mathrm{popcount}(\cdot)$  counts the number of  `1's. For real-valued  HVs (INT8 or FP32), cosine similarity is instead applied~\cite{kang2023homstc,annsolo}. In the binary setting, the XOR + popcount formulation aligns well with IMP, NMP, and ISP platforms that support simple, massively parallel operation for large-scale OMS workloads~\cite{imani2017voicehd,kang2022massively}. In the remainder of this work, we refer to this query-reference similarity scoring operation, e.g., XOR + popcount for binary HDC, or cosine similarity for INT8 and FP32, as the \textit{similarity kernel}, since it is the dominant computing primitive repeatedly executed across all evaluated platforms.

%% file: TEX/3_4_5.Architectures_Accelerators_Mapping.tex
\section{Cross-Domain OMS Accelerator Architectures}
\label{sec:arch-mapping}

This section summarizes representative OMS accelerators based on various platforms, starting with 1) commodity hardware such as GPU and FPGA, and then moving to 2) memory-centric platforms that use DRAM, PCM, ReRAM, NAND, and FeNAND for IMP, NMP, and ISP execution.


For each platform, we describe the accelerator by focusing on (1) dataflow, indicating where the reference library is placed, e.g., on-device memory, near-storage buffer, or in-array storage, (2) how the similarity kernel (FP32 or INT8 dot-product or binary Hamming distance calculation) is processed, and (3) how the selected top-k outputs are reduced and returned to the host. The methodology for the cross-platform comparisons is defined separately in Section~\ref{sec:oms-scaling}.

\subsection{OMS acceleration on commodity hardware}\label{sec:commodity}


\begin{figure}[h]
    \centering
    \includegraphics[width=0.5\textwidth]{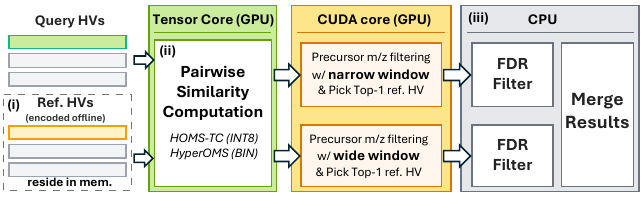}
    \caption{GPU mapping of HDC-based OMS similarity scoring and top-$k$ reduction~\cite{kang2023homstc}.}
    \label{fig:gpu_arch}
\end{figure}

\noindent \textbf{1) GPU (NVIDIA RTX 4090 and H100)}:
As described in Section~\ref{sec:background}, OMS execution follows a common flow across platforms: a similarity kernel scores queries against the reference library, and the resulting top-$k$ candidates are reduced and passed to downstream FDR filtering. Fig.~\ref{fig:gpu_arch} shows the GPU instantiation of this flow, where (i) reference shards reside in GPU device memory, (ii) similarity scoring is executed on the GPU using tensor-core INT8 or binary Hamming distance kernels, and (iii) compact top-$k$ outputs are transferred to the CPU for merge and subsequent FDR processing.

GPUs offer an effective solution for OMS acceleration by combining high device-memory bandwidth (GPU $\leftrightarrow$ GDDR/HBM)  with massive SIMT (single instruction, multiple threads) parallelism, where many lightweight threads execute the same instruction sequence on different data for the similarity computation with tensor cores as shown in Fig.~\ref{fig:gpu_arch}, where 24~GB GDDR6X is employed on RTX~4090 and 80~GB HBM on H100. There are two representative GPU implementations: (1) ANN-SoLo~\cite{annsolo}, which performs FP32 cosine similarity search, and (2) HDC-based pipelines such as HyperOMS~\cite{kang2022massively} and HOMS-TC~\cite{kang2023homstc}, which simplifies the similarity kernel to XOR + popcount (binary) or INT8 tensor-core operations, substantially improving throughput and energy efficiency. 
When the reference library exceeds the capacity of a single GPU, it is partitioned into shards that are processed sequentially or distributed across multiple devices. GPUDirect Storage~\cite{nvidia-gpudirect-storage}, which enables direct data transfer from NVMe storage to GPU memory without staging through host DRAM, reduces redundant data movement by allowing reference shards to stream directly into the GPU. Per-shard top-$k$ results are then merged either on the GPU or on the host CPU to form the final ranked list for downstream FDR filtering.




\noindent \textbf{2) FPGA (e.g., SmartSSD~\cite{smartssd21}):}

\begin{figure}[h]
    \centering
    \includegraphics[width=.48\textwidth]{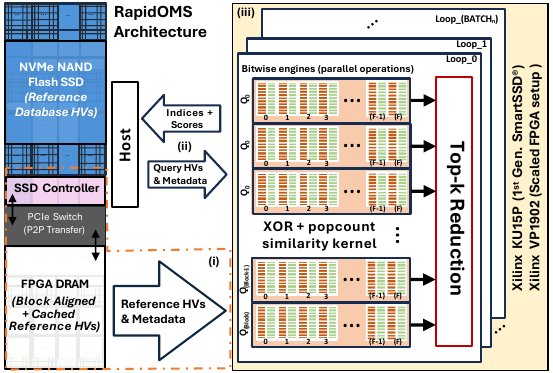}
    \caption{SmartSSD~\cite{smartssd21} OMS: data flow between near-storage FPGA with device DRAM and SSD controller.}
    \label{fig:fpga_arch}
\end{figure}

Near-storage FPGA systems integrate reconfigurable logic together with FPGA-local DRAM within the SSD module, forming a computational storage device that allows reference traffic to bypass the host (Fig.~\ref{fig:fpga_arch}). Commercial SmartSSD devices~\cite{smartssd21}, for example, place a Kintex UltraScale+ KU15P FPGA together with its DRAM buffer at the SSD controller position, connecting it directly to the NAND flash array and the NVMe interface. Systems that co-locate a newer VP1902 FPGA\cite{amd2023vp1902} with NVMe storage provide approximately $16\times$ more logic and on-board memory resources in a similar near-storage configuration, enabling wider parallelism and deeper buffering while preserving the same streaming dataflow.

RapidOMS~\cite{pinge2024rapidoms} maps OMS to FPGA-friendly bitwise engines organized as a 3-stage streaming pipeline (Fig.~\ref{fig:fpga_arch}): (i) reference HV blocks are staged from NAND into FPGA-local DRAM and then streamed into the bitwise engines, (ii) query HVs and metadata are streamed from the host and broadcast/partitioned across parallel bitwise-engine lanes, and (iii) the FPGA evaluates the XOR + popcount similarity kernel and maintains a local top-$k$, returning only compact indices and scores to the host. When the working set exhibits reuse, frequently accessed reference blocks remain cached in FPGA DRAM across query batches; otherwise, the SSD controller continues to refill and stream blocks while computation proceeds on the current block. RapidOMS further exploits FPGA-side control for fine-grained, metadata-driven filtering (e.g., precursor-window gating) to avoid unnecessary comparisons, and SmartSSD-style peer-to-peer connectivity can redistribute reference data directly between devices without staging through host DRAM.

\subsection{Memory-Centric OMS acceleration}\label{sec: mem-centric arch}

\begin{figure}[h]
    \centering
    \includegraphics[width=0.48\textwidth]{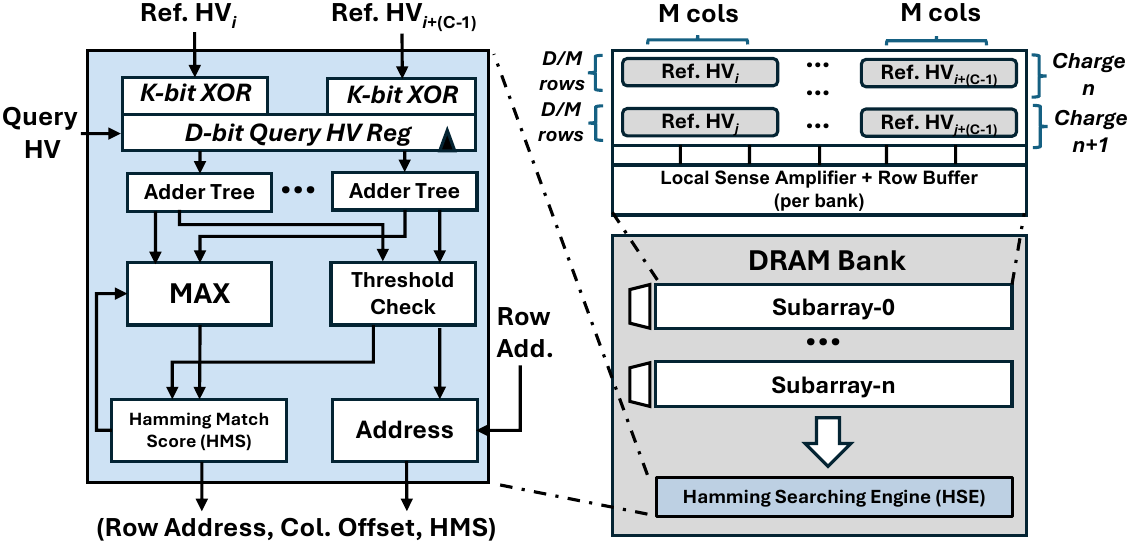}
    \caption{DRAM near-memory OMS~\cite{kang2024dram} with near-bank Hamming Searching Engines (HSE).}
    \label{fig:dram_IMP}
\end{figure}
\noindent \textbf{1) DRAM-Centric NMP Accelerator:} 
Unlike commodity accelerators that transfer reference shards from memory into a separate compute hierarchy, DRAM-centric near-memory processing (NMP) executes the similarity kernel directly at the memory banks where reference HVs reside~\cite{kang2024dram}. As a result, DRAM channels are used mainly for broadcasting queries and returning compact result outputs, rather than repeatedly transferring the full reference corpus.

%

As illustrated in Fig.~\ref{fig:dram_IMP}, lightweight near-bank Hamming Searching Engines (HSEs) are placed adjacent to the local sense amplifiers and row buffers of each DRAM bank. Reference HVs are stored as fixed-length $M$-bit slices, with each activated row for $C$ slices in disjoint column regions, separated by fixed column offsets, enabling parallel evaluation of multiple candidates per row. Across the full HV dimensionality $D$, each reference is scanned over $\frac{D}{M}$ successive row activations.
During execution, a query HV is latched into a per-bank register, streamed as $M$-bit slices, and broadcast across banks. On each row activation, the HSE performs an $M$-bit XOR between the query slice and all reference slices, placed with fixed column offsets, accumulating partial Hamming distances via a local popcount adder tree as the scan progresses.

To reduce wasted work, the engine employs a successive search procedure: it first evaluates a subset of bits to form a coarse score, prunes low-quality candidates, and completes the full evaluation only for survivors. This early-exit mechanism eliminates a substantial fraction (50-75\%)  of bit operations with negligible impact on identification quality, improving effective throughput by concentrating compute on likely matches. Each bank maintains a local top list during scanning, while the dispatcher aggregates bank-local results into a global top-$k$ (or ranked list).

References are interleaved across banks by $m/z$ range so that adjacent $m/z$ intervals are distributed over different banks, which helps mitigate skew and load imbalance. The dispatcher maps reference IDs to $(\text{bank}, \text{row range}, \text{column offset})$, issues the corresponding activation sequences, and performs the final reduction. Because similarity is computed in-place, off-chip traffic is dominated by query broadcasts and compact outputs rather than reference movement, in contrast to GPU-based shard streaming. The design scales naturally by provisioning additional DRAM modules, increasing both capacity and bank-level parallelism.

\noindent \textbf{2) ReRAM and PCM-Centric IMP Accelerators:}
Unlike DRAM-centric OMS, where similarity evaluation is performed digitally near the row buffer, ReRAM-~\cite{fan2024mlcrram} and PCM-based~\cite{fan2024specpcm} IMP accelerators map reference HVs directly into analog conductances within crossbar arrays and perform dot-product operations entirely in memory. By computing similarity in place, these designs eliminate repeated streaming of reference shards through the memory hierarchy and expose massive array-level parallelism, shifting the primary system bottleneck from external memory bandwidth to peripheral throughput, most notably DAC/ADC conversion and accumulation. Both ReRAM and PCM support multi-level-cell (MLC) operation (3--4 bits per cell), enabling dense storage of reference HVs and efficient in-memory analog similarity computation.

\begin{figure}[h]
\centering
\includegraphics[width=0.37\textwidth]{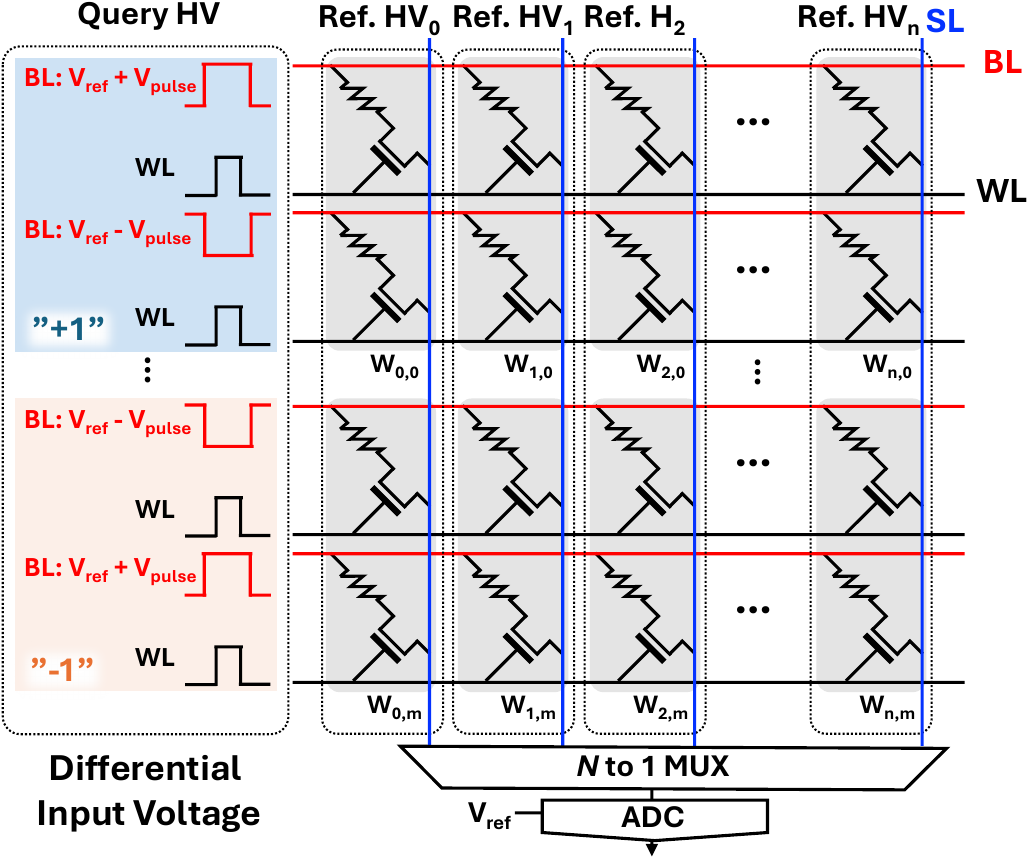}
\caption{ReRAM-based IMP for OMS~\cite{fan2024mlcrram}: in-memory VMM operation with ADC readout.}
\label{fig:reram}
\end{figure}


As shown in Fig.~\ref{fig:reram}, ReRAM-based IMP OMS accelerators~\cite{fan2024mlcrram} encode reference HVs as conductance values in 1T1R crossbar arrays. The reference HVs are stored in each column, and each column outputs the vector's similarity score. During vector-matrix multiplication (VMM), the input query HV is simultaneously transmitted through differential bitline (BL) voltages to represent signed inputs. All $N$ activated columns contributing to the MAC output receive a high signal from the wordline (WL), and the resulting currents are collected by the capacitor, generating a voltage on the sourceline (SL) according to the equation $V_{SL} = \frac{\sum I \cdot \Delta t}{C_{SL}}$ in parallel. 
Then, ADC  digitizes the analog SL voltages, after which near-memory digital logic accumulates or ranks the partial similarity scores.
While ReRAM offers high storage density (approximately $3\times$ higher than high-density SRAM), MLC ReRAM exhibits notable non-idealities, including conductance relaxation (10--30\% drift within the first hour after programming), high write voltages (3--5~V), cycle-to-cycle variation (8--15\%), and relatively low resistance on/off ratios (often $<50\times$ for MLC), which reduce the analog signal margin and make similarity computation more susceptible to noise.  To mitigate these effects, ReRAM-based designs employ differential weight encoding and leverage the inherent error tolerance of HD computing, which can sustain up to $\sim$10\% memory errors without significant accuracy loss. 

\begin{figure}[h]
\centering
\includegraphics[width=0.40\textwidth]{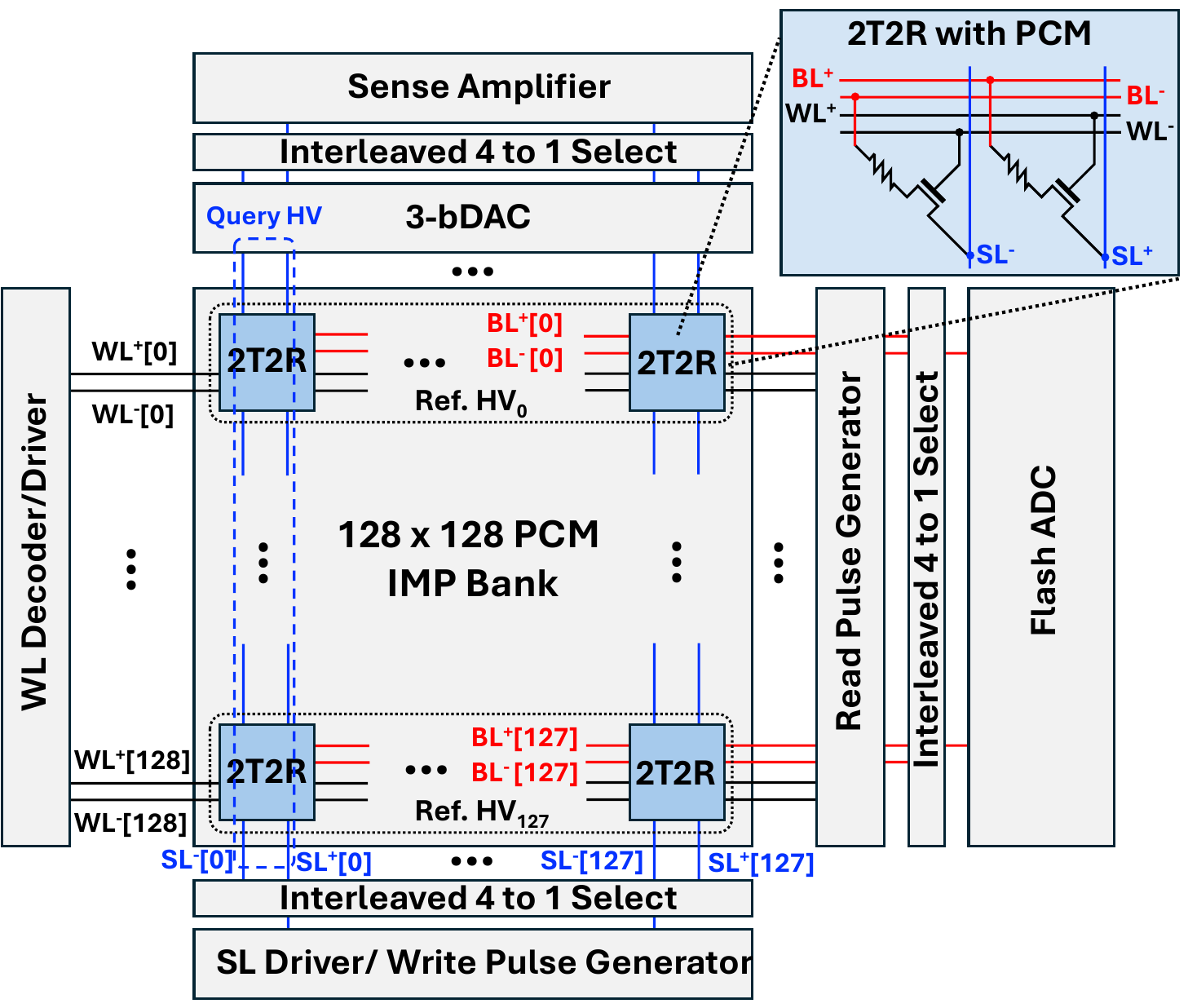}
\caption{PCM-based IMP for OMS~\cite{fan2024specpcm}: crossbar arrays with DAC/ADC for MLCs.}
\label{fig:pcm}
\end{figure}


The PCM-based IMP OMS accelerator~\cite{fan2024specpcm}, illustrated in Fig.~\ref{fig:pcm}, follows a similar analog VMM principle where reference HVs are stored as conductance states in two consecutive devices in a row using 2T2R differential cells (i.e., signed values are represented by the conductance difference between the paired cells). On top of the sign representation, the 2T2R configuration suppresses MLC drift by canceling common-mode conductance variation between paired cells.  During VMM, the query HV is converted by a 3-bit DAC and applied as analog input voltages on the SLs, while WLs are concurrently enabled to broadcast the input across multiple rows for parallel dot-product evaluation. The resulting analog partial sums appear on the differential BLs (BL$^{+}$/BL$^{-}$) and are digitized by shared flash ADCs, after which near-memory logic performs ranking/selection. By employing PCM devices based on superlattice-material concepts (e.g., using $Ge_{4}Sb_{6}Te_{7}$-based stacks)~\cite{wu2024novel}, these designs reduce programming voltage to below 1.0~V and achieve resistance on/off ratios of 100$\times$--150$\times$, providing substantially improved linearity and noise margins compared to ReRAM. 

While both ReRAM and PCM support MLC operation, HDC inherently exploits binary weights, which underutilizes the available storage density when a purely binary representation is used. To address this limitation, \cite{fan2024specpcm}
proposes \emph{dimension packing}, where multiple binary HV dimensions are compressed into a single MLC value by summing the binary values of $n$ consecutive dimensions, effectively increasing storage and compute density by a factor of $n$.
For example, when $n=3$ and each binary dimension has a value of $\{-1,+1\}$, three consecutive dimensions can be aggregated into a packed decimal value. This produces four possible sums:
$(+1,+1,+1)\rightarrow +3$,
$(+1,+1,-1)\rightarrow +1$,
$(+1,-1,-1)\rightarrow -1$, and
$(-1,-1,-1)\rightarrow -3$.
Accordingly, three binary dimensions can be represented by four logical levels $\{-3,-1,+1,+3\}$, which are mapped onto the available conductance states of the MLC cell.
Although the resulting dot-product values are not exactly identical to those obtained from the original binary-based distance computation, they largely preserve the relative distances among HVs due to the high dimensionality and sparsity of HVs, thereby maintaining the similarity ranking between HVs.
During search, these packed values are used directly for in-memory dot-product evaluation, reducing the effective HV dimensionality from $D$ to $D/n$ while increasing storage and compute density. In this work, we apply dimension packing to both devices under the assumed 3-bit/cell MLC configurations, thereby improving effective storage density and parallel search throughput.
%
At the same time, MLC operation introduces a marginal bit error rate (BER) due to reduced noise margins and device-level variability. The impact of this BER on OMS accuracy is
quantitatively evaluated in Section~\ref{sec:acc_preservation}, with representative BER statistics summarized in Table~\ref{tab:ber_imp}.

The throughput of IMP-based OMS accelerators is primarily determined by the interplay between array level parallelism, the read delay of the array during IMP operations, and ADC bandwidth. Each array typically comprises 128 to 256 rows or columns per bank, enabling highly parallel operations by accessing these cells at a time. The evaluation is performed across multiple banks or over multiple cycles to accommodate long HVs, which commonly have around 8k dimensions.
ADC bandwidth is governed not only by the intrinsic speed of the ADC itself, but also by the degree of sharing across multiple rows or columns to satisfy area constraints. The shared ADCs digitize the analog output in a time-interleaved way, causing potential bandwidth limitation.
Prior work explores different design points along this spectrum. For example, the ReRAM-based design in~\cite{fan2024mlcrram} adopts aggressive ADC sharing across tens to hundreds of columns to amortize the high area and power cost of high-speed ADCs, particularly when the ADC operates significantly faster than the array read. In contrast, the PCM-based implementation in~\cite{fan2024specpcm} employs fewer, higher precision flash ADCs shared across a smaller number of rows when the ADC is relatively slow and sufficient area budget for ADCs is available. These examples highlight how ADC sharing ratios can be tuned based on device characteristics, area budget, and target throughput.

\subsection{Storage-Centric Accelerators (3D NAND and FeNAND)}

\begin{figure}[h]
    \centering
    \includegraphics[width=0.5\textwidth]{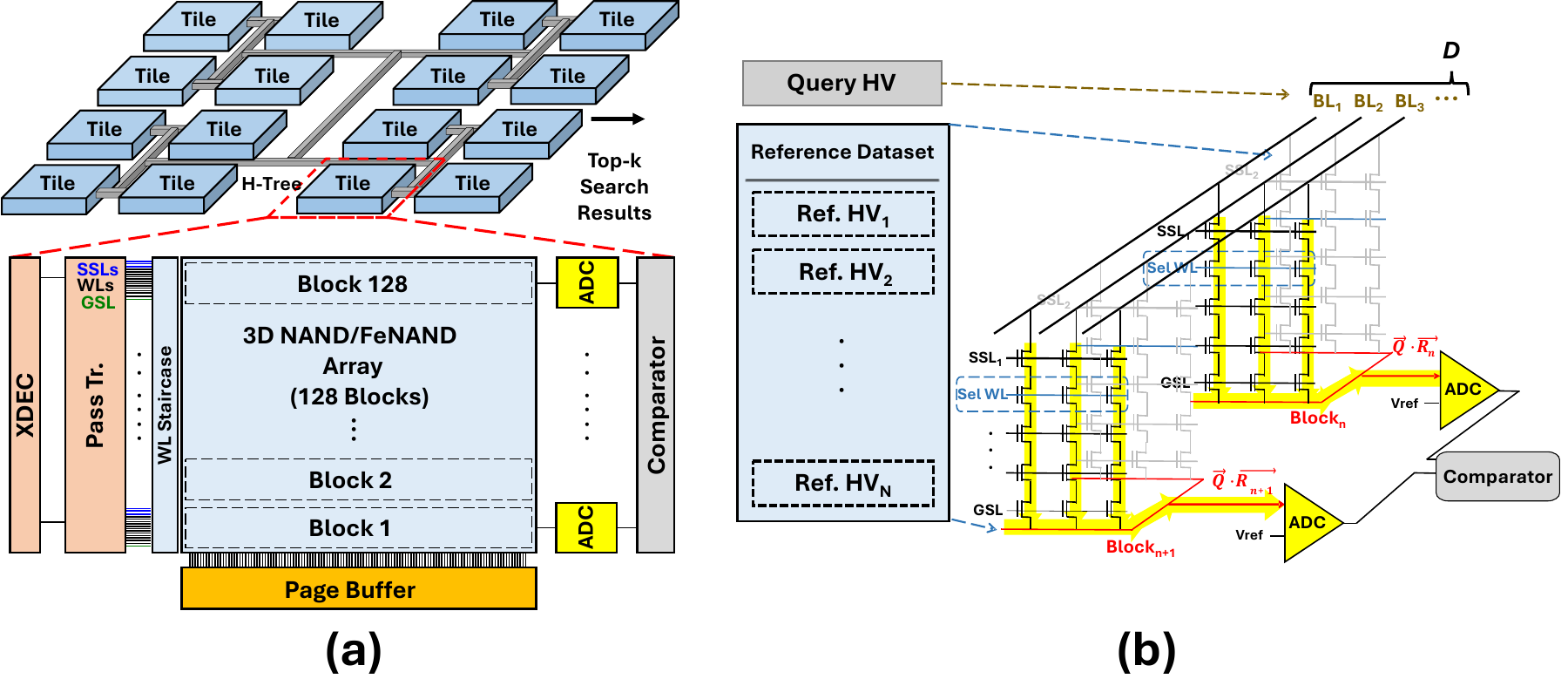}
    \caption{In-storage processing (ISP) on 3D NAND / FeNAND for OMS~\cite{hsu2025fenand}: (a) reconfigurable multi-tile architecture that combines tiles via an on-package H-tree network for query broadcast and result reduction, and ISP tile showing 3D NAND array organization, and (b) reference HV mapping and sensing path. }
    \label{fig:nand_fenand}
\end{figure}

DRAM-, ReRAM-, and PCM-based NMP and IMP solutions move the similarity evaluation kernel into the memory platform, which eliminates memory-to-host staging of reference data. 
However, at repository scale, proteomics corpora are orders of magnitude larger than practical DRAM capacity and are typically stored on SSD-backed infrastructure. For example, the PRIDE Archive reports datasets totaling 2.85 PB as of Aug. 2024~\cite{pride2025update}, and other public proteomics repositories exceed 900 TB \cite{massive2024} in stored data. In this scenario, even with high SSD bandwidth, repeatedly moving the data from the SSD  to main memory dominates both latency and energy. In-storage processing (ISP)~\cite{hsu2023memsys} addresses this bottleneck by executing similarity evaluation in-place inside the SSD package so that references are not staged out of the storage device.


Fig.~\ref{fig:nand_fenand} describes the overview of the 3D NAND/FeNAND ISP structure. A key architectural distinction from ReRAM/PCM crossbars is that NAND flash does not support crossbar-style analog accumulation across many activated wordlines. On the other hand, cells on a NAND array are connected \emph{serially} along a string, preventing the  Kirchhoff current summation used for VMM by enabling multiple WLs at a time in the crossbar array. As a result, NAND-based ISP performs \emph{page-parallel} evaluation, where all the cells connected to a single WL are executed at a time,  with lightweight mixed-signal sensing and digital ranking, rather than a fully parallel across the array. However, such an operation can be performed from multiple tiles in parallel. An on-package H-tree network enables query broadcast and result reduction across multiple tiles.

Over the past decade, 3D NAND has scaled from tens of layers to close to 300 layers, but further $z$-scaling is increasingly limited by reliability challenges (e.g., cell-to-cell interference) and high write voltages in charge-trap nitride (CTN) based 3D NAND~\cite{1000layers}. FeNAND replaces the CTN layer with a ferroelectric layer whose spontaneous polarization enables lower-voltage operation and provides a pathway for continued vertical scaling and higher areal density (reported projections beyond 1000 layers and $>100$\,Gb/mm$^2$)~\cite{lim2023comprehensive,das2023experimental,1000layers}. The lower array operating voltage improves the energy efficiency for both the cell programming and sensing operations.

A 3D NAND block comprises a three-dimensional grid of vertical channels (strings) and horizontal WLs: each WL--string intersection forms a transistor cell, and multiple cells in a string are connected in series between a BL at the top and a ground-select line (GSL) at the bottom. String-select lines (SSLs) and the GSL gate the ends of each string to enable selective activation during read and program operations, while BLs connect to sense amplifiers (SAs) and associated page buffers at the array periphery. During a read, all BLs are sensed in parallel, while a single WL (and typically one SSL group) is enabled at a time. This organization exposes high parallelism at the page level, even though cells along a string are serially connected.

An ISP tile connects a 3D NAND/FeNAND array with lightweight mixed-signal and digital periphery, as illustrated in Fig.~\ref{fig:nand_fenand}(a). References are stored as binary HVs: within each tile, a reference HV is mapped onto a single page by assigning each HV bit to a dedicated BL. Following Hsu~et~al.~\cite{hsu2023memsys,hsu2025fenand}, the page width (number of BLs connected to a WL) is chosen to be at least the HV dimension $D$, so a single page access provides the full $D$-bit reference HV without additional partial passes.
During search, the controller applies the query HV as a set of binary BL voltage patterns and selects the WL corresponding to a given reference. This page-parallel access evaluates thousands of BLs simultaneously; the resulting source-line currents represent the accumulated overlap between the query and the selected reference (i.e., a Hamming-style similarity). Tile-local page buffers and ADCs digitize these currents to produce per-reference similarity scores; a shared comparator tree then ranks scores across blocks within the tile to select the tile-local top-$k$ candidates. 
Because similarity scoring is performed in-place inside the ISP tiles, host traffic remains small and only weakly dependent on library size; effective throughput is primarily governed by the number of active tiles per package and the achievable concurrency across pages, planes, and dies.

%% file: TEX/5.OMS_Model.tex
\section{Methodology for Comparative Study}\label{sec:eval}

Given the wide variety of device characteristics, memory capacities, and operating mechanisms across accelerator platforms, we adopt a consistent evaluation setup across datasets, algorithms, and hardware configurations. This section also summarizes how we reproduce and scale platforms from the prior works and define the evaluated workloads.

\subsection{Unified Algorithm and Workload Configuration}
\label{sec:algo-workload}
Our study evaluates both conventional FP32 OMS and HDC-based OMS across GPUs, near-storage FPGAs, and memory-centric accelerators. 
On commodity GPUs, we consider ANN-SoLo, a widely used, state-of-the-art open modification spectral-library search (OMS) implementation. ANN-SoLo reports the strongest identification yield compared with established peptide-search tools~\cite{annsolo,lam2007spectrast,kong2017msfragger} by performing FP32 shifted dot-product, or cosine, similarity using vector units.
We therefore use ANN-SoLo as the primary conventional OMS software baseline and directly evaluate it on a GPU (RTX~4090) under the shared HEK293 setup for the speedup and energy-efficiency comparisons. We also test ANN-SoLo on a CPU (Intel i7-8700K with 64\,GB DDR4) under the same conditions as an additional reference. In addition, we evaluate HOMS-TC (INT8 HDC) and HyperOMS (binary HDC) on the GPU platforms. Because GPUs natively support FP32, INT8, and binary arithmetic, these implementations serve as software OMS baselines.

In contrast, in- and near-memory processing on multi-level non-volatile memories (e.g., ReRAM, PCM, 3D NAND, and FeNAND) is constrained by limited sensing margins and device variability under MLC operation, making high-precision FP32 or INT8 dot-products impractical for these devices. For these platforms, we therefore adopt the HyperOMS formulation, where spectra are encoded as binary HVs and similarity is evaluated using massively parallel in-array VMM operations or bitwise XOR + popcount primitives implemented near the memory periphery. We use HyperOMS in this binary HD form as the common OMS kernel for DRAM-NMP, ReRAM/PCM IMP, and 3D NAND/FeNAND ISP, enabling a numerically appropriate and portable mapping across these memory-centric platforms.


We evaluate mid- and large-scale workloads as follows.

\noindent  $\bullet$ \textbf{Mid-scale dataset.} The \emph{mid-scale} configuration targets a single-accelerator setting and uses HEK293 (Human Embryonic Kidney 293) as the query dataset, consisting of multiple query files (b1906--b1938) with an average of 46{,}665 query spectra per file. Each query spectrum is embedded as an $D=$ 8{,}192-dimensional vector and searched against the MassIVE-KB human spectral reference library with $N = 2{,}992{,}672$ references. The same library is encoded in three forms for GPU execution: \(m/z\)-intensity peaks for ANN-SoLo (about 10\,GB), INT8 HDC vectors for HOMS-TC (about 22.8\,GB), and binary HDC vectors for HyperOMS (about 2.85\,GB). 

\noindent  $\bullet$ \textbf{Large-scale dataset.} 
To capture \emph{large-scale} OMS behavior, where the reference library exceeds the capacity of a single accelerator, we construct a terabyte-scale binary HDC workload that preserves the same query contents as the mid-scale setting (8{,}192-D HVs) and uses 46{,}665 query spectra, while scaling the reference library to \(N = 10^{9}\) reference spectra. This corresponds to approximately 1\,TB of reference data in binary form (and about 8\,TB in INT8), which exceeds single-accelerator memory capacity and therefore requires GPU-based execution with consistent streaming of the reference library.
On the other hand, ISP platforms retain the library persistently within SSD-scale 3D NAND and FeNAND packages.
Based on these two workload scales, we organize the evaluation using a three-tier methodology that includes (i) baseline algorithmic efficiency on GPUs, (ii) platform-dependent execution efficiency for the mid-scale deployment, and (iii) large-scale deployment at the TB limit, where the reference library no longer fits in an accelerator’s local memory.


Table~\ref{tab:hw_summary} summarizes the mapping of workloads, algorithms, and scale for the evaluations for each hardware platform. The  GPU-based software execution is performed to identify the speedup and energy gains from the software variations, such as HD-based kernels and reduced precision. 
The mid-scale evaluation was performed on all the hardware platforms. 
On the other hand, the large-scale evaluation restricts evaluation to NVIDIA H100 and full SSD-scale ISP architectures (3D NAND and FeNAND), excluding DRAM-NMP and ReRAM/PCM-IMP due to impractical capacity scaling, and highlights the contrast between repeatedly loading reference shards over PCIe/NVLink versus executing similarity evaluation in-place within ISP devices at terabyte-scale. 


\begin{table}[t]
\caption{OMS workloads and evaluated device--algorithm configurations. ``Mid-scale'' uses HEK293 queries against the MassIVE-KB reference library; ``Large-scale'' corresponds to the 1\,TB binary HDC reference library ($N{=}10^9$, $D{=}8192$).}
\label{tab:hw_summary}
\centering
\scriptsize
\setlength{\tabcolsep}{6.5pt}
\renewcommand{\arraystretch}{1.0}
\begin{tabularx}{\columnwidth}{c c c}
\toprule
\textbf{Device / class} & \textbf{Algorithm} &  \textbf{Dataset}  \\
\midrule
CPU (Intel i7-8700K) & ANN-SoLo (FP32 cosine) & Mid-scale \\
\midrule
\multirow{3}{*}{GPU (RTX~4090)} &
ANN-SoLo (FP32 cosine) & Mid-scale \\
& HOMS-TC (INT8 HDC)    & Mid-scale \\
& HyperOMS (Binary HDC)  & Mid-scale \\
\midrule
\multirow{3}{*}{GPU (H100)} &
ANN-SoLo   & Mid-scale \\
& HOMS-TC     & Mid-scale \& Large-scale \\
& HyperOMS   & Mid-scale \& Large-scale \\
\midrule
Near-storage FPGA          & HyperOMS & Mid-scale \\
DRAM NMP                   & HyperOMS & Mid-scale \\
ReRAM IMP                  & HyperOMS & Mid-scale \\
PCM IMP                    & HyperOMS & Mid-scale \\
3D NAND ISP                & HyperOMS & Mid-scale \& Large-scale \\
FeNAND ISP                 & HyperOMS & Mid-scale \& Large-scale \\
\bottomrule
\end{tabularx}
\end{table}


\begin{table*}[t]
\centering
\caption{Hardware parameters for various OMS acceleration platforms.}
\label{tab:params_platform}
\small
\begin{tabular}{l l l l l l l}
\hline
\textbf{Platform} & \textbf{Memory  device} & \textbf{Capacity} & \textbf{Search primitive} & \textbf{Peri. circuitry} & \textbf{I/O / staging path} \\
\hline
GPU (RTX 4090) &
GDDR6X & 24\,GB &
CUDA/Tensor Cores &
-- &
Host$\rightarrow$GPU \\
GPU (H100) &
HBM3 & 80\,GB &
CUDA/Tensor Cores &
-- &
Host$\rightarrow$GPU \\
\hline
FPGA (VP1902)~\cite{pinge2024rapidoms} &
On-board DRAM &
8~GB &
XOR + popcount &
-- &
NAND$\rightarrow$DRAM refill \\
\hline
DRAM NMP~\cite{kang2024dram} & 
DDR4 &
8~GB &
XOR + popcount &
-- &
Host$\rightarrow$DRAM provisioning\\
\hline
ReRAM IMP~\cite{fan2024mlcrram} &
In-array (MLC)&
3~GB &
Analog VMM + ADC &
7-b SAR ADC &
Pre-programmed refs\\
PCM IMP~\cite{fan2024specpcm} &
In-array (MLC) &
3~GB &
Analog VMM + ADC &
6-b Flash ADC; 3-b DAC&
Pre-programmed refs \\
\hline
Storage ISP~\cite{hsu2023memsys,hsu2025fenand} &
In-storage arrays &
1~TB &
In-storage scoring &
6-b SAR ADC &
Pre-programmed refs \\
\hline
\end{tabular}
\end{table*}

\subsection{OMS Workload Behavior}\label{sec:oms-scaling}

Other than the spectrum encoding phase, OMS performance is governed by library size, representation precision, and whether the reference library fits within the executing device’s local memory. Although arithmetic cost scales with vector dimension $D$ and the number of reference vectors evaluated per query, the end-to-end throughput is often dominated by the cost of the reference data movement when it must be loaded from a larger storage tier.

In more detail, the data movement behavior is determined by how large the reference library is, as compared to the device’s memory.
If the device memory is sufficient, the reference library (or working set) fits locally and can be reused across query batches. In this case, performance is primarily determined by in-device memory bandwidth and the efficiency of the similarity primitive computation. Throughout the remainder of this manuscript, this regime is referred to as the \emph{search-only} delay.
If the device memory is insufficient, the system repeatedly streams reference blocks from off-chip memory or storage. Since each block must be loaded into device memory once before the similarity search is performed, this regime incurs additional data transfer overhead. We denote this case as the \emph{search + $1\times$write/load} delay and analyze it in the following sections.

%% file: TEX/6.MergedEval.tex
\definecolor{resultpurple}{HTML}{000000}
\newcommand{\result}[1]{\textcolor{resultpurple}{\bfseries #1}}
  
\section{Experimental Results}\label{Results}



\subsection{Experimental Setup}
We evaluate OMS accelerators by following the OMS and methodology from Section~\ref{sec:eval}. 
We report throughput and energy for both \emph{search-only} and \emph{search + $1\times$ load}. 
This separation isolates kernel processing efficiency from the staging overhead required to stream data consistently. 
The detailed experiment setup for each device is described below.

For the results reported in Tables~\ref{tab:hek-managed},~\ref{tab:hek-hyperoms-only}, and ~\ref{tab:hek-large}, we use a unified evaluation flow that combines our measured software baselines and estimations following the evaluation methodology of prior works  under the same workload. Specifically, we evaluated software baseline ANN-SoLo on CPU (i7-8700K) and GPU (RTX~4090), respectively; the H100 and VP1902 entries are projected from measured RTX~4090 and SmartSSD/RapidOMS baselines using the scaling methodology described below; and the DRAM-NMP, ReRAM, PCM, 3D NAND, and FeNAND entries are evaluated using the latency/energy models, device parameters, peripheral assumptions, and dataflows from the corresponding prior works.

\noindent $\bullet$ \textbf{CPU baseline.}
We use ANN-SoLo running on an Intel i7-8700K with 64\,GB DDR4 as the conventional CPU baseline for the mid-scale HEK293 workload. This baseline provides a conventional software reference point in Tables~\ref{tab:hek-managed} and~\ref{tab:hek-hyperoms-only}.



\noindent $\bullet$ \textbf{GPUs.}
We measure throughput and energy on an NVIDIA RTX~4090 (GDDR6X, 24\,GB, $\sim$1\,TB/s) using three kernels: ANN-SoLo (FP32 cosine), HOMS-TC (INT8 tensor-core HDC), and HyperOMS (binary HDC). The results of NVIDIA H100 (HBM3, 80\,GB, $\sim$3.35\,TB/s) are estimated using a compute throughput- and data bandwidth-based scaling. Specifically, search-kernel latency is scaled from the RTX~4090 baseline using the ratio of effective low-precision throughput, where H100 (SXM) provides 3958\,TFLOPS FP8 tensor throughput while RTX~4090 provides 1321.22\,TFLOPS FP8-equivalent throughput, yielding a scaling factor of \(3958/1321.22=2.996\times\). This estimate is also consistent with the memory-subsystem difference, since the H100 HBM3 bandwidth is approximately \(3.35\times\) higher than that of  RTX~4090 GDDR6X. Therefore, both compute-throughput and memory-bandwidth ratios are close to \(3\times\), supporting the projected H100 results. Energy is estimated from the scaled latency and nominal board power, using 450\,W for RTX~4090 and 700\,W for H100 SXM.  When the data volume exceeds the GPU memory capacity, the GPU streams shards through \emph{GPUDirect Storage (GDS)}.

While we execute the GPU baselines (ANN-SoLo, HOMS-TC, HyperOMS) on an RTX~4090, we reproduce the results of the remaining platforms using the reported latency/energy models and dataflow described in the corresponding works, e.g., DRAM timing, ADC sharing, INP/ISP/NMP delay parameters, and parallel or sequential access patterns~\cite{kang2022massively,kang2023homstc,annsolo,pinge2024rapidoms,fan2024specpcm,kang2024dram,fan2024mlcrram,hsu2023memsys,hsu2025fenand}.

\noindent $\bullet$ \textbf{Near-storage FPGAs.}
For the FPGA evaluations, we measure search latency and power on a Samsung/Xilinx SmartSSD CSD (Kintex UltraScale+) and estimate results for a Xilinx Versal VP1902 (co-located with NVMe storage and on-board DRAM) by scaling the SmartSSD measurements to reflect the VP1902’s higher logic and on-board buffering resources (Section~\ref{sec:commodity}) under the same dataflow. 
Specifically, search latency is scaled based on the relative FPGA logic resources. The VM1802-class scaling baseline referenced in~\cite{pinge2024rapidoms} provides 899K LUTs, whereas the VP1902 provides 8460K LUTs, yielding a projected kernel-scaling factor of \(8460/899=9.41\times\). This is motivated by the design’s throughput being primarily determined by replicated bitwise XOR + popcount lanes with parametric parallelism, together with metadata-driven pruning to reduce unnecessary comparisons. For the DRAM access delay, the write/load component was modeled separately from the search-computation latency, which was scaled according to the LUT-capacity ratio. We assumed the same DDR4-2400 memory subsystem for both FPGAs, corresponding to a theoretical peak bandwidth of 19.2~GB/s per 64-bit channel. For the SSD access delay, we used the sequential write bandwidth of a high-performance PCIe~5.0 NVMe SSD~\cite{WD} for both FPGAs. In summary, the VP1902 performance projection from the VM1802 baseline separately accounts for the search-computation latency scaled by the LUT-capacity ratio, while applying the same DRAM and SSD write/load access delays.
For energy, we use a conservative active-power scaling model. 
Since the VP1902 projection instantiates approximately \(9.41\times\) more parallel search lanes, 
we approximate the active  search power to scale with the degree of replicated parallelism from the 40\,W VM1802-class baseline.
 Combined with the latency scaling above, this yields approximately constant search energy while reflecting the trade-off between higher throughput and higher active power.

\noindent $\bullet$ \textbf{DRAM-NMP.} 
For DRAM NMP, we employ commodity DDR4-2400 modules and estimate near-bank processing delay/energy based on Kang~et~al.~\cite{kang2024dram}, while write energy follows Kim~et~al.~\cite{kim2013memory}. The baseline configuration employs 8\,GB of DDR4-2400 DRAM (16 chips $\rightarrow$ 256 banks) to hold the mid-scale reference library. 
Near-bank search/compute is modeled as repeated ACT$\rightarrow$compute$\rightarrow$PRE over DRAM rows, using the DDR4-2400 timing parameters from~\cite{kang2024dram} (e.g., $t_{CK}{=}0.833$\,ns and $t_{RC}{=}55$ cycles); throughput is determined by the internal bank/sub-array bandwidth and per-bank DRAM timing, while the external DDR4 I/O channel is not on the critical path since it carries only query broadcast and compact top-$k$ outputs.


\noindent  $\bullet$ \textbf{ReRAM/PCM IMP.}
For ReRAM and PCM IMP designs, we adopt an in-array storage scenario, i.e., the reference HVs are stored directly within the crossbar arrays and reused for similarity computation across query batches. 
The PCM-based design~\cite{fan2024specpcm} is modeled at its native 40\,nm technology node as reported in prior work. In contrast, since prior ReRAM IMP prototypes~\cite{fan2024mlcrram} are reported at a 130\,nm technology node, we scale the ReRAM device to a 40\,nm node for comparative evaluation by following established scaling methodologies~\cite{stillmaker2017scaling}. We note that this scaling introduces limitations because the methodology is developed for digital circuits, while the ReRAM IMP architecture exhibits mixed signal behavior. 

We model the array capacity to be sufficient to store the mid-scale reference library in the IMP fabric by using multiple banks. Under this target, both devices require the same number of physical cells as both utilize 3-bit/cell MLC (or triple-level cell, TLC) and differential pairing (two cells per logical weight) for signed representation.
The detailed architecture follows \cite{fan2024mlcrram} for ReRAM, while the PCM follows \textit{SpecPCM}~\cite{fan2024specpcm} as introduced in Section~\ref{sec: mem-centric arch}.
Critically, we adopt the peripheral and ADC configurations from the baselines~\cite{fan2024mlcrram, fan2024specpcm} as they are highly optimized with the inherent device characteristics. Therefore, ReRAM utilizes a high-ratio column-sharing SAR ADC (e.g., 128--256:1) to mitigate area penalties. In contrast, PCM employs an 8-row-sharing flash ADC, a configuration enabled by the high read-bandwidth from the array and sensing stability of superlattice devices.

\begin{table}[t]
\centering
\caption{Effective BER  for MLC IMP devices.}
\label{tab:ber_imp}
\small                   
\renewcommand{\arraystretch}{1.1}
\setlength{\tabcolsep}{6pt}      
\begin{tabular}{lcc}
\toprule
\textbf{Device} & \textbf{MLC Configuration} & \textbf{Effective BER} \\
\midrule
ReRAM IMP~\cite{fan2024mlcrram} & 3-bit/cell (TLC) & 7.65\% \\
PCM IMP~\cite{fan2024specpcm}   & 3-bit/cell (TLC) & 4.5\% \\
\bottomrule
\end{tabular}
\end{table}



\noindent $\bullet$  \textbf{3D NAND / 3D FeNAND ISP.}
For 3D NAND and FeNAND ISP designs, we adopt device parameters, latency and energy models, and the ISP execution flow reported by Hsu et al.~\cite{hsu2025fenand}. We include peripheral overheads, including sense paths, ADCs, and controller logic, in both latency and energy. FeNAND follows the same ISP array organization and controller schedule as 3D NAND, but uses a ferroelectric transistor-based memory cell that is compatible with mature 3D NAND processing. This cell replaces the charge trapping layer with a ferroelectric layer, such as Hf--Zr--O~\cite{hsu2025fenand}. Consistent with the fabricated device characteristics reported by Hsu et al.~\cite{hsu2025fenand}, we employ a ferroelectric layer thickness of approximately 10\,nm and a thin interfacial oxide of about 8\,\AA, which together enable lower voltage operation than the conventional 3D NAND. In our modeling, this primarily reduces memory access energy in the array and sensing paths, while leaving the page and tile-parallel concurrency unchanged.

Table~\ref{tab:params_platform} summarizes the evaluated platforms, per-package capacities, and the search primitives used in each mapping.
 
\vspace{-1.1mm}

\subsection{OMS Identification Quality (Accuracy)}
\label{sec:acc_preservation}

\begin{figure}[t]
  \centering
  \includegraphics[width=0.97\columnwidth]{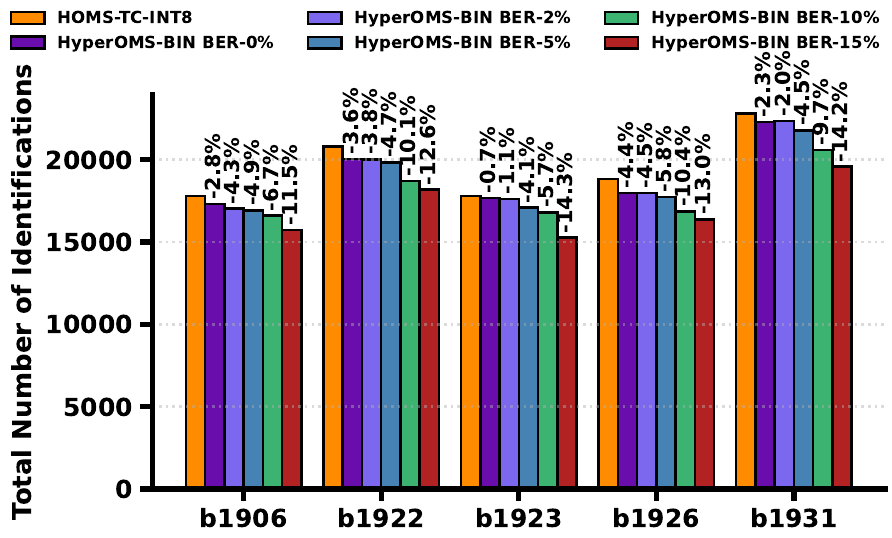}
  \caption{Impact of injected BER on HyperOMS identifications at fixed FDR for datasets b1906--b1931. Bars report total identifications for HOMS-TC (INT8) and HyperOMS (BIN) under injected BER.}
  \label{fig:hyperoms-ber}
\end{figure}

\begin{figure*}[t]
  \centering
  \includegraphics[width=\textwidth]{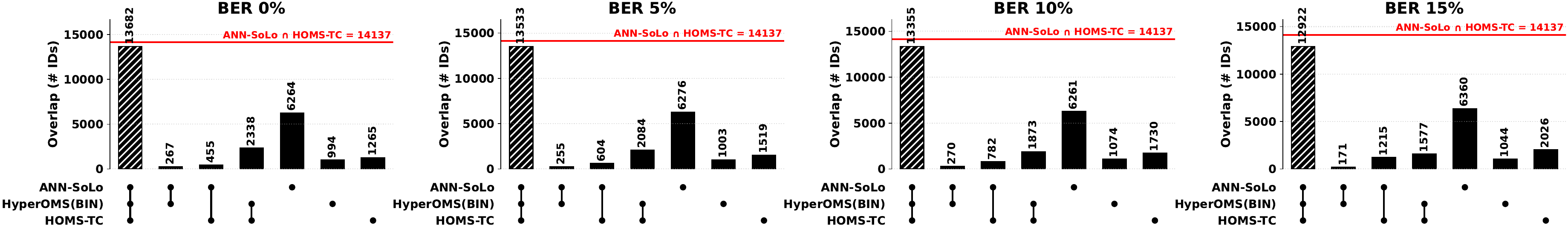}%
  \caption{Identification consistency across algorithms (UpSet intersections, b1906). Each graph corresponds to one BER setting; bars report the number of intersections among ANN-SoLo, HOMS-TC, and HyperOMS.}
  \label{fig:UPSET}
\end{figure*}



\noindent $\bullet$  \textbf{Analog Non-ideality Modeling.} We quantify how device-level uncertainty affects identification accuracy in-memory centric OMS execution. To emulate non-ideal behavior, we inject random bit flips into the binary HVs given a controlled BER. This bit flip model captures read errors that arise when multi-level ReRAM and PCM cells operate near their sensing limits.
We sweep BER values based on reported error characteristics of multi-level ReRAM and PCM arrays, as shown in Fig.~\ref{fig:hyperoms-ber}, and use these ranges as representative operating conditions for IMP execution. The specific BER settings used in our experiments are summarized in Table~\ref{tab:ber_imp}, indicating 4 - 7\% BER from the analog ReRAM and PCM IMP with a 3-bit MLC configuration. Unless otherwise noted, we treat the remaining evaluated platforms as effectively error-free at the bit level, since they rely on single-level binary storage whose raw BER is negligible compared to the injected ranges considered here.

We evaluate OMS identification quality by widely employed two complementary methods as described below.

\noindent $\bullet$  \textbf{Total identifications.} 
First, we measure coverage via total peptide-spectrum identifications at an FDR threshold of 1\%, which is standard practice in OMS evaluations~\cite{Goloborodko2013Pyteomics,annsolo,kang2023homstc,kang2022massively}. Across datasets b1906--b1931, HyperOMS tracks the HOMS-TC baseline closely, given the BER $\le 10\%$. At 15\% BER, the reduction in total identifications remains modest and dataset dependent ($\sim$9--14\%), indicating that discovery coverage degrades smoothly as the BER increases. 
Note that Fig.~\ref{fig:hyperoms-ber} reports the total number of identifications passing a fixed FDR threshold, which reflects coverage. As a result, it does not directly reflect the quality or consistency of the identifications. For this reason, we additionally employ UpSet overlap analysis, as described below.

\noindent $\bullet$  \textbf{UpSet overlap.} 
An UpSet plot (a set-intersection visualization) in Fig.~\ref{fig:UPSET} quantifies how the identified set changes across three different methods: 1) ANN-SoLo, 2) HyperOMS, and 3) HOMS-TC, to indicate the robustness. 

Because ground truth is not available for HEK293, we treat the consensus between
ANN-SoLo and HOMS-TC as a proxy reference. This intersection captures
identifications supported by both a conventional scoring baseline and an 8-bit HDC baseline, and therefore serves as a conservative approximation of
high-confidence peptides for robustness comparison. We then quantify $\mathrm{Recovery}$ as the fraction of this proxy consensus that is also
identified by HyperOMS:

{\footnotesize
\begin{equation}
\mathrm{Recovery} \;=\; 
\frac{|\mathrm{ANN}\_\mathrm{SoLo} \cap \mathrm{HOMS\text{-}TC} \cap \mathrm{HyperOMS}\_\mathrm{BIN}|}
{|\mathrm{ANN}\_\mathrm{SoLo} \cap \mathrm{HOMS\text{-}TC}|}
\end{equation}
}
This metric indicates how many queries yield the same result as the proxy reference. It corresponds to the first bar in each graph of Fig.~\ref{fig:UPSET}, where multiple graphs are shown for BER values of $\{0,5,10,15\%\}$ in consideration of 4 - 8\% BER from the analog ReRAM and PCM IMP as shown in Table~\ref{tab:ber_imp}. 
The remaining bars indicate overlaps between different methods. For example, the second bar shows the number of queries for which ANN-SoLo and HyperOMS produce the same identification.


Given the UpSet plot, $\mathrm{Recovery}$ is $96.78\%$ at $0\%$ BER, $95.73\%$ at $5\%$ BER, $94.47\%$ at $10\%$ BER, and $91.41\%$ at $15\%$ BER. These results show that identification consistency remains high up to $10\%$ BER, while $15\%$ BER marks the onset of more noticeable, yet still moderate, degradation.
Total identifications and UpSet overlaps indicate that HyperOMS maintains substantial discovery coverage and preserves a large fraction of high-confidence peptides under realistic device-level BER, supporting its suitability for memory-centric deployments.
Taken together, total identification counts and UpSet overlap analysis indicate that HyperOMS maintains substantial discovery coverage and preserves a large fraction of high-confidence peptides. This supports its suitability for memory-centric deployments using XOR + popcount or in-array VMM primitives under realistic device-level BER.

\begin{table*}[t]
  \centering
  \captionsetup{labelfont=normalfont,textfont={color=black}}
  \caption{Normalized energy efficiency and speedup for algorithmic comparisons over RTX 4090 (ANN-SoLo) baseline with mid-scale dataset (HEK293)}
  \label{tab:hek-managed}

  \setlength{\fboxsep}{4pt}
  \setlength{\fboxrule}{0.8pt}

    \begin{minipage}{0.965\textwidth}
      \centering
      \begin{tabularx}{\textwidth}{l *{7}{Y}}
        \toprule
        Device & {i7-8700K} & {RTX 4090} & {RTX 4090} & {RTX 4090} & {H100} & {H100} & {H100} \\
        Memory & {DDR4} & {GDDR6X} & {GDDR6X} & {GDDR6X} & {HBM3} & {HBM3} & {HBM3} \\
        Algorithm & {ANN-SoLo} & {ANN-SoLo} & {HOMS-TC} & {HyperOMS} & {ANN-SoLo} & {HOMS-TC} & {HyperOMS} \\
        Data Type & {FP32} & {FP32} & {INT8} & {BIN} & {FP32} & {INT8} & {BIN} \\
        \midrule
        Energy effic. (search-only) &
        \result{0.53} &
        \result{1.00} & \result{9.69} & \result{18.81} &
        \result{1.93} & \result{18.71} & \result{37.25} \\
        Energy effic. (search + 1$\times$ write/load) &
        \result{0.53} &
        \result{1.00} & \result{9.67} & \result{18.80} &
        \result{1.93} & \result{18.61} & \result{37.21} \\
        \midrule
        Speedup (search-only) &
        \result{0.79} &
        \result{1.00} & \result{23.71} & \result{33.78} &
        \result{2.99} & \result{71.03} & \result{101.19} \\
        Speedup (search + 1$\times$ write/load) &
        \result{0.80} &
        \result{1.00} & \result{20.51} & \result{32.94} &
        \result{2.97} & \result{48.40} & \result{93.62} \\
        \bottomrule
      \end{tabularx}
    \end{minipage}%
\end{table*}

\begin{table*}[t]
  \centering
  \captionsetup{labelfont=normalfont,textfont={color=black}}
  \caption{Normalized energy efficiency and speedup across platforms over RTX 4090 (HyperOMS) baseline with mid-scale dataset (HEK293)}
  \label{tab:hek-hyperoms-only}

  \setlength{\fboxsep}{4pt}
  \setlength{\fboxrule}{0.8pt}

    \begin{minipage}{0.965\textwidth}
      \centering
      \begin{tabularx}{\textwidth}{l *{9}{Y}}
        \toprule
        Platform / Device & {i7-8700K} & {RTX 4090} & {H100} & {VP1902} & {DRAM} & {ReRAM} & {PCM} & {3D NAND} & {FeNAND} \\
        Method / Type & {ANN-SoLo} & {GPU HDC} & {GPU HDC} & {Near-Storage} & {NMP} & {IMP} & {IMP} & {ISP} & {ISP} \\
        Data Type & {FP32} & {BIN} & {BIN} & {BIN} & {BIN} & {BIN} & {BIN} & {BIN} & {BIN} \\
        \midrule
        Energy effic. (search-only) &
        \result{0.03} & \result{1.00} & \result{1.98} & \result{2.84} &
        \result{65.54} & \result{6,737.43} & \result{16,187.91} & \result{4,912.42} & \result{40,743.24} \\
        Energy effic. (search + 1$\times$ write/load) &
        \result{0.03} & \result{1.00} & \result{1.97} & \result{2.84} &
        \result{63.92} & \result{6,741.88} & \result{16,198.61} & \result{4,915.67} & \result{40,770.16} \\
        \midrule
        Speedup (search-only) &
        \result{0.02} & \result{1.00} & \result{2.99} & \result{3.09} &
        \result{1.87} & \result{2.50} & \result{23.10} & \result{17.01} & \result{17.01} \\
        Speedup (search + 1$\times$ write/load) &
        \result{0.02} & \result{1.00} & \result{2.84} & \result{2.76} &
        \result{1.76} & \result{2.57} & \result{23.75} & \result{17.49} & \result{17.49} \\
        \bottomrule
      \end{tabularx}
    \end{minipage}%
\end{table*}

\vspace{-1.1mm}

\subsection{Energy and Delay Comparisons}
\label{sec:comparative_results}

\noindent $\bullet$  \textbf{Algorithmic Efficiency on GPUs.} 
Having established that HyperOMS preserves identifications under realistic device-level BER (Section~\ref{sec:acc_preservation}), we next isolate the algorithmic benefit of reduced-precision operations on commodity GPUs. 
Table~\ref{tab:hek-managed} compares CPU and GPU ANN-SoLo with HOMS-TC and HyperOMS on RTX~4090 and H100, with all results normalized to RTX~4090 ANN-SoLo.
The CPU ANN-SoLo result provides a conventional software reference, achieving $0.79\times$ search throughput and $0.53\times$  energy efficiency as compared to RTX~4090 ANN-SoLo in the search-only case.
On RTX~4090, reducing precision from FP32 to INT8/BIN HDC yields substantial benefits.
HOMS-TC achieves a $23.71\times$ speedup and $9.69\times$ energy-efficiency improvement in the search-only case, while HyperOMS further increases these to $33.78\times$ and $18.81\times$, respectively.
Scaling to H100 amplifies the same trend: relative to the RTX~4090 FP32 baseline, ANN-SoLo achieves $2.99\times$ speedup and $1.93\times$ energy-efficiency improvement, whereas HOMS-TC and HyperOMS reach $71.03\times$ / $18.71\times$ and $101.19\times$ / $37.25\times$, respectively.
The performance gap between RTX~4090 and H100 primarily stems from H100's higher memory bandwidth, greater on-chip parallelism, and improved efficiency for low-precision workloads. 
Including the one-time write/load cost leaves energy-efficiency trends largely unchanged, but significantly reduces speedup for H100 HOMS-TC, e.g., the speedup drops sharply from $71.03\times$ to $48.40\times$.
This reduction occurs because the write/load overhead becomes significant relative to the RTX~4090 ANN-SoLo baseline, where FP32 computation dominates execution time and data movement overhead is less pronounced.
In contrast, for H100 HyperOMS, the impact of write/load overhead is smaller due to reduced data precision. As a result, the speedup decreases more modestly, from $101.19\times$ to $93.62\times$, since binary representations substantially reduce the volume of data transferred.

Overall, these results indicate that the improvement comes from changing the number representation and its underlying compute primitive, moving from FP32 cosine similarity to reduced-precision HDC and ultimately to a bitwise kernel.

\begin{table}[t]
\caption{Kernel-level efficiency comparison over various platforms under the mid-scale workload. (search-only)}
\label{tab:generalized_search_only}
\centering
\footnotesize
\setlength{\fboxsep}{0.1pt} 
\setlength{\fboxrule}{0.8pt} 
\begin{tabular}{lcc}
\toprule
Platform & Throughput (comp./s) & Energy Effic. (comp./J) \\
\midrule
i7-8700K (ANN-SoLo)   & $4.49 \times 10^{8}$  & $1.63 \times 10^{6}$  \\
RTX 4090 (ANN-SoLo)   & $5.66 \times 10^{8}$  & $3.08 \times 10^{6}$  \\
RTX 4090 (HOMS-TC)    & $1.34 \times 10^{10}$ & $2.98 \times 10^{7}$  \\
RTX 4090 (HyperOMS)   & $1.91 \times 10^{10}$ & $5.79 \times 10^{7}$  \\
H100 (ANN-SoLo)       & $1.70 \times 10^{9}$  & $5.94 \times 10^{6}$  \\
H100 (HOMS-TC)        & $4.02 \times 10^{10}$ & $5.76 \times 10^{7}$  \\
H100 (HyperOMS)       & $5.73 \times 10^{10}$ & $1.15 \times 10^{8}$  \\
FPGA (VP1902)           & $5.92 \times 10^{10}$ & $1.65 \times 10^{8}$  \\
DRAM NMP              & $3.58 \times 10^{10}$ & $3.79 \times 10^{9}$  \\
ReRAM IMP             & $4.79 \times 10^{10}$ & $3.90 \times 10^{11}$ \\
PCM IMP               & $4.42 \times 10^{11}$ & $9.37 \times 10^{11}$ \\
3D NAND ISP           & $3.26 \times 10^{11}$ & $2.84 \times 10^{11}$ \\
FeNAND ISP            & $3.26 \times 10^{11}$ & $2.36 \times 10^{12}$ \\
\bottomrule
\end{tabular}
\end{table}

\begin{table*}[t]
  \centering
  \caption{Normalized energy efficiency and speedup across platforms over H100 (HOMS-TC) baseline with large-scale dataset.}
  \label{tab:hek-large}
  \adjustbox{max width=\textwidth}{
  \begin{tabularx}{\textwidth}{l *{4}{Y}}
    \toprule
    Platform / Device & {H100} & {H100} & {3D NAND} & {FeNAND} \\
    Method & {GPU HDC} & {GPU HDC} & {ISP} & {ISP} \\
    Data Type & {INT8 (HOMS-TC)} & {BIN (HyperOMS)} & {BIN} & {BIN} \\
    \midrule
    Energy efficiency (search-only) &
    \result{1.00} & \result{1.99} & \result{4,938.36} & \result{40,958.38} \\
    Energy efficiency (search + 1$\times$ write/load) &
    \result{1.00} & \result{2.00} & \result{4,966.98} & \result{41,187.48} \\
    \midrule
    Speedup (search-only) &
    \result{1.00} & \result{1.42} & \result{8.09} & \result{8.09} \\
    Speedup (search + 1$\times$ write/load) &
    \result{1.00} & \result{1.96} & \result{12.19} & \result{12.19} \\
    \bottomrule
  \end{tabularx}}
\end{table*}

\noindent $\bullet$  \textbf{Mid-scale cross-platform comparisons.} 
This comparison evaluates the accelerator platforms by using the same algorithm, HyperOMS with binary HDC, across all accelerators, thereby removing algorithmic differences and allowing the hardware platforms to be compared on a consistent basis.
Results in Table~\ref{tab:hek-hyperoms-only} are normalized to RTX~4090 running HyperOMS. The CPU ANN-SoLo column provides a conventional FP32 software reference, achieving only $0.02\times$  search throughput and $0.03\times$ energy efficiency as compared to RTX~4090 HyperOMS in the search-only case.


Relative to the RTX~4090 HyperOMS baseline, H100 improves throughput by $2.99\times$ and energy efficiency by $1.98\times$, while a VP1902-class near-storage FPGA reaches $3.09\times$ speedup and $2.84\times$ energy efficiency in the search-only case. These results indicate that, in the mid-scale case, moving from a GPU implementation to a carefully designed FPGA accelerator improves efficiency by customizing the architecture for the specific operation, but does not fundamentally address the data-movement cost.

In contrast, memory- and storage-centric designs achieve orders-of-magnitude energy gains by avoiding repeated data movement of the reference library from the host to the processor by executing similarity where the data is stored. DRAM-NMP improves energy efficiency to $65.54\times$ while maintaining comparable throughput by shifting the dominant cost from off-chip transfers to near-bank/sub-array operations. Non-volatile IMP designs (ReRAM and PCM) execute similarity computation directly at the array/periphery level and therefore remove bulky data movement, reaching energy-efficiency gains of $6.74 \times 10^{3}$ times for ReRAM and $1.62 \times 10^{4}$ times for PCM. In addition, PCM achieves a $23.10\times$ speedup, while ReRAM is more constrained at $2.50\times$ due to ADC sharing across columns, which limits effective throughput.

NAND-based ISP extends this principle to storage-class memory, yielding energy-efficiency gains from $4.91 \times 10^{3}$ times for 3D NAND to $4.07 \times 10^{4}$ times for FeNAND. These architectures achieve $17.01\times$ speedup by leveraging massive BL parallelism, processing 8,192-dimensional vectors in a single cycle. FeNAND attains higher energy efficiency due to differences in the underlying device technology. Contemporary high-density 3D NAND based on charge-trap transistors is mature and cost-effective for storing massive spectral datasets, but its high operating voltages, often exceeding 30\,V, result in high energy consumption. In contrast, ferroelectric transistors enable FeNAND operation at much lower voltages, typically below 5\,V, which reduces energy per access in the array and sensing paths and therefore improves overall energy efficiency at comparable throughput~\cite{hsu2025fenand}. The identical speedups for 3D NAND and FeNAND occur because they share a unified ISP microarchitecture where throughput is constrained by digital logic overheads, such as H-tree based query broadcasting and serial top-k sorting, rather than by the raw access speed of the memory cells. However, both 3D NAND and FeNAND achieve lower speedups than PCM due to sequential WL enabling, whereas PCM enables all WLs in parallel.

Finally, the search-only versus search+$1\times$write/load rows remain close for most platforms in this table because the working reference library fits in device memory and can be reused across multiple queries, allowing the one-time load and write cost to be amortized.

To provide an absolute view of kernel-level efficiency under the same OMS workload, Table~\ref{tab:generalized_search_only} reports the search-only throughput and energy efficiency for the  mid-scale setting in terms of vector-level comparisons per second (comp./s) and comparisons per joule (comp./J), where one comparison denotes one query-reference's vector-to-vector similarity evaluation. The results show a clear contrast: commodity GPU implementations remain in the $10^{9}$--$10^{10}$ comparisons/s regime, FPGA, DRAM NMP, and ReRAM IMP reach around $10^{10}$ comparisons/s, and PCM IMP and NAND ISP accelerators reach the $10^{11}$ comparisons/s range. The separation is even more pronounced in energy efficiency, where GPU baselines stay around $10^{6}$--$10^{8}$ comparisons/J, DRAM NMP reaches $10^{9}$ comparisons/J, and non-volatile IMP/ISP platforms span $10^{11}$--$10^{12}$ comparisons/J. These absolute metrics are consistent with the normalized trends in Table~\ref{tab:hek-hyperoms-only} and highlight that the major gain of memory-/storage-centric designs comes not only from parallelism in similarity evaluation, but also from minimizing repeated data movement of the reference library.

\noindent $\bullet$  \textbf{Large-scale cross-platform comparisons.} 
The large-scale comparison evaluates system-level scalability when the reference library exceeds the memory capacity of the accelerator. Table~\ref{tab:hek-large} increases the binary library size to $N=10^9$ ($\approx$1\,TB for $D{=}8192$ BIN) and normalizes results to H100 running INT8 HOMS-TC, reporting both search-only and search+$1\times$ write/load metrics.


For the search-only case, where the reference working set already resides in GPU memory, the BIN HDC kernel on H100 (HyperOMS) provides only a modest advantage over
INT8, improving energy efficiency by $1.99\times$ and speed by $1.42\times$.
Once the practical requirement of fetching the reference library into device memory is considered in the search + $1\times$ write/load case, the GPU must repeatedly reload reference shards from storage. In this regime, the benefit of the BIN representation becomes more pronounced, with speedup increasing from $1.42\times$ to $1.96\times$, as the smaller bit precision reduces the volume of data transferred across the external interface. Despite this, the corresponding energy efficiency gain changes only marginally because total energy is dominated by GPU-side kernel execution rather than data movement from SSD to GPU.

Storage-centric ISP avoids this data movement by executing similarity where the repository resides.  
In Table~\ref{tab:hek-large}, the normalized speedup from search-only ($8.09\times$) to search+$1\times$ write/load ($12.19\times$) reflects the fact that the GPU baseline pays a large reference fetching cost at this scale, whereas ISP does not introduce such a penalty. Both 3D NAND and FeNAND show the same speedups as shown in the mid-scale dataset, given their identical microarchitecture and throughput bottleneck from the digital logic.  
However, the ferroelectric transistor-based cell technology primarily improves the energy per operation.
Relative to H100 INT8, FeNAND achieves $\sim4.12\times10^4$ times energy savings, while 3D NAND achieves $\sim4.97\times10^3$ times because only compact results traverse the host interface.
Overall, Table~\ref{tab:hek-large} highlights the increased benefit of storage-centric ISP over H100 compared to Table~\ref{tab:hek-hyperoms-only}, as GPU execution becomes increasingly bottlenecked by external data movement.


%% file: TEX/7.Discussions.tex
\section{Discussions}
\label{sec:discussion}
This work uses OMS as a representative, memory-intensive similarity-based search workload to study architectural and cross-device trade-offs that extend beyond proteomics.
A key insight from this study is that binary HDC enables a common OMS kernel across heterogeneous platforms. By reducing similarity evaluation to XOR + popcount or in-array VMM primitives, HyperOMS naturally maps to GPUs, as well as memory-centric IMP and ISP, where high-precision MAC support is impractical.
While emerging devices combined with analog operation introduce marginal BER due to reduced noise margins and device-level variability, HDC inherently tolerates this noise because HVs are separated by large distances in high-dimensional space. As a result, moderate bit perturbations do not significantly alter similarity ordering. This tolerance enables aggressive MLC usage and dimension packing, which directly translates into higher storage density and higher effective throughput.

Scalability in such memory-intensive applications is governed by reference data movement rather than peak compute capability. When the reference library can be reused without repeated write/load operations from external memory, performance is determined by internal bandwidth and parallelism. As the dataset grows, repeated staging of reference blocks through external interfaces increasingly dominates execution time and energy. This behavior explains why storage centric ISP remains well matched to repository-scale OMS.

This paper places less emphasis on the non-volatile nature of emerging devices, yet this property can be highly beneficial for mobile systems~\cite{chang2011circuit, chiu2010low, 11074827}, where devices remain idle for a large fraction of time, dramatically reducing idle power consumption. In addition, this work focuses on data that can be deployed on SSDs even for large-scale benchmarks, whereas systems at the scale of hundreds of TBs may require more systematic connectivity and communication, potentially across numerous storage devices. In such regimes, advanced algorithms that dynamically narrow the comparison set and selectively fetch the relevant data can be synergistic with search-centric applications.

While this work focuses on OMS, the same principles underlying the strong benefits of ISP and IMP extend to a broader class of search-centric and data-intensive workloads, such as vector database algorithms~\cite{ma2023comprehensive, jing2025large} that increasingly rely on approximate search methods and accept probabilistic correctness metrics such as recall rate. To leverage emerging devices for such workloads, controlled error arising from device non-ideality can be traded for substantial gains in performance and efficiency. Accordingly, modeling device non-ideality and predicting system-level accuracy under controlled error conditions are emphasized as essential components for guiding system design and reliable deployment of such devices.

%% file: TEX/8.Conclusions.tex
\section{Conclusion}
\label{sec:Conclusion}

This paper introduces and compares a unified, workload-driven evaluation of OMS accelerators across commodity processors and emerging memory and storage centric platforms, including GPUs, near-storage FPGAs, DRAM NMP, ReRAM and PCM IMP, and 3D NAND and FeNAND ISP. By evaluating these platforms under consistent algorithmic and accuracy assumptions, our results indicate that colocating similarity computation with data—through IMP, NMP, and especially ISP—eliminates repeated reference staging and yields substantial gains in energy efficiency and throughput. In addition, binary HDC provides a common similarity kernel across these heterogeneous platforms. It maps efficiently to commodity GPUs using bitwise primitives, while its inherent error tolerance enables robust execution on memory-centric architectures despite realistic device level non-idealities. Overall, this study highlights the importance of in-situ processing near the data  as the primary driver of acceleration for OMS and related similarity-based search workloads, motivating memory and storage centric architectures as a practical path toward energy- and delay-efficient large-scale search acceleration.

\section*{Acknowledgment}
This work was supported in part by the Center for Processing with Intelligent Storage and Memory (PRISM) under Semiconductor Research Corporation (SRC) grant 2023-JU-3135, and CoCoSys, both centers in JUMP 2.0, an SRC program sponsored by DARPA. This work was also supported in part by NSF grants \#2112665, \#2211386, and \#2112167.